\newif\ifIEEE
\IEEEtrue

\ifIEEE 

\documentclass[lettersize,journal]{IEEEtran}
\usepackage{amsmath,amsfonts}
\usepackage{algorithmicx,algpseudocode}
\usepackage{array}
\usepackage[caption=false,font=normalsize,labelfont=sf,textfont=sf]{subfig}
\usepackage{textcomp}
\usepackage{stfloats}
\usepackage{url}
\usepackage{verbatim}
\usepackage{graphicx}
\def\BibTeX{{\rm B\kern-.05em{\sc i\kern-.025em b}\kern-.08em
    T\kern-.1667em\lower.7ex\hbox{E}\kern-.125emX}}
\usepackage{balance}
\usepackage{xspace}

\usepackage{tabularx}
\usepackage{array}
\usepackage{booktabs}
\usepackage[table]{xcolor}

\else 

\documentclass{article}
\usepackage{graphicx} 
\usepackage{fullpage}
\usepackage{amsfonts,amsmath,amsthm}
\usepackage{algorithmicx, algorithm, algpseudocode}
\usepackage{xspace}
\usepackage{framed}
\usepackage{booktabs}
\usepackage[font={small}]{caption}
\usepackage{stmaryrd}

\fi   

\usepackage{xcolor}
\usepackage{tikz}
\usetikzlibrary{decorations.pathreplacing}

\usepackage{hyperref}

\renewcommand{\paragraph}[1]{\medskip\noindent\textbf{#1}}

\newcommand{\N}{\mathbb{N}}     
\newcommand{\Z}{\mathbb{Z}}     
\newcommand{\F}{\mathbb{F}}     
\newcommand{\G}{\mathbb{G}} 

\newtheorem{theorem}{Theorem}[section]

\newtheorem{definition}[theorem]{Definition}

\newtheorem{remark}[theorem]{Remark}

\newcommand{\zo}{\{0,1\}\xspace}

\newcommand{\Eval}{{\sf Eval}}
\newcommand{\eval}{{\sf Eval}}
\newcommand{\Gen}{{\sf Gen}}
\newcommand{\gen}{{\sf Gen}}
\newcommand{\Convert}{{\sf Convert}}
\newcommand{\Keep}{{\sf Keep}}
\newcommand{\Lose}{{\sf Lose}}

\newcommand{\Sim}{{\sf Sim}}

\newcommand{\Real}{{\sf Real}}
\newcommand{\Ideal}{{\sf Ideal}}

\newcommand{\ynote}[1]{\color{blue}[Ynote: #1]\color{black}}

\newcommand{\cF}{\mathcal{F}}
\newcommand{\bbG}{\mathbb{G}}
\newcommand{\rin}{r^{{\sf in}}}
\newcommand{\rout}{r^{{\sf out}}}

\newcommand{\secp}{\lambda}

\newcommand{\bs}{\mathbf{s}}

\newcommand{\gin}{{\sf in}}
\newcommand{\gout}{{\sf out}}

\newcommand{\fulleval}{{\sf FullEval}}
\newcommand{\distgen}{{\sf DistGen}}
\newcommand{\leak}{{\sf Leak}}

\newcommand{\Leak}{{\sf Leak}}

\newcommand{\ceil}[1]{\left\lceil #1 \right\rceil}

\newcommand{\pt}{\bullet}

\title{Distributed Point Functions and Function Secret Sharing}

\author{
    Elette Boyle (\IEEEmembership{Senior Member, IEEE}), 
    Niv Gilboa, 
    Yuval Ishai, 
    and Peter Scholl (\IEEEmembership{Member, IEEE})%
\thanks{E.\ Boyle is with the Cryptography and Information Security (CIS) Lab at NTT Research, Sunnyvale (e-mail: eboyle@alum.mit.edu).}
\thanks{N.\ Gilboa is with the Faculty of Computer and Information Sciences, Ben Gurion University, Be'er Sheva, Israel (e-mail: gilboan@bgu.ac.il).}
\thanks{Y.\ Ishai is with the Department of Computer Science, Technion Institute of Technology, Haifa, Israel, currently on a sabbatical at AWS, New York. Supported by ISF grant 3527/24 and BSF grant 2022370. This article is not associated with Amazon. (e-mail: yuvali@cs.technion.ac.il).}
\thanks{P.\ Scholl is with the Department of Computer Science, Aarhus University, Aarhus, Denmark (e-mail: peter.scholl@cs.au.dk).}
}

\date{}

\begin{document}

\maketitle

\begin{abstract}
A {\em distributed point function} (DPF) is a cryptographic primitive that enables compressed additive sharing of a secret weight-1 vector (equivalently, a point function)
across two or more parties. The appealing lightweight structure of DPF constructions has enabled a wide range of applications. These include private information retrieval, anonymous messaging, secure computation with preprocessing, and {\em pseudorandom correlation generators} for expanding small correlated seeds into large pseudorandom instances of cryptographic correlations.
    
In this article, we survey definitions, constructions, and applications of DPFs. We also discuss the extension of DPF to {\em function secret sharing} (FSS), which generalizes point functions to support richer function classes. Efficient FSS schemes yield a similar generalization for most of the applications of DPFs.
\end{abstract}

\section{Introduction}




Secret sharing is a central tool in cryptography. In its simplest form, a secret-sharing scheme splits a secret $s$ into a random pair of shares $(s_0,s_1)$ such that $s_0+s_1=s$. This kind of {\em additive} secret sharing is typically done over a finite Abelian group $\G$. In this case, each share perfectly hides $s\in\G$ if we pick $s_0$ uniformly at random from $\G$ and let $s_1=s-s_0$. Considering additive sharing over $\G^N$,  we can similarly split a secret {\em vector} $\bs\in\G^N$ into a pair of uniformly random share vectors $(\bs_0,\bs_1)$ such that $\bs=\bs_0+\bs_1$.

Can we compress the secret shares $\bs_0,\bs_1$ of a long vector $\bs$ while still ensuring that each individual share hide $\bs$? If we insist on {\em perfect} hiding, then the above solution can be easily shown to be optimal: namely, each share must be uniformly random and therefore cannot be compressed. However, if we settle for computational hiding and assume the existence of a pseudorandom generator (PRG), or equivalently a one-way function~\cite{HILL99}, then a limited amount of compression becomes possible. Letting $r$ be a seed for a PRG $G$ that stretches a random $\lambda$-bit seed $r$ to a pseudorandom element of $\G^N$, where $\lambda \ll N$, we can split $\bs\in\G^N$ into $(r, \bs - G(r))$. By locally expanding the first share to $G(r)$, we obtain additive secret-sharing of $\bs$ with {\em computational} (rather than perfect) secrecy. This can make the first share much shorter than $\bs$, but (inevitably) yields at most a 2x reduction in the {\em total} share size, since the two shares together should allow recovering $\bs$. For general $\bs$, the required description size---and thus combined secret share size---must necessarily be at least $N \lceil\log |\G|\rceil$ bits.

\paragraph{Function secret sharing (FSS).} The barrier for compressing the shares of a general long vector $\bs$ is its large description size. But what if $\bs$ is taken from a small set of vectors, so that $\bs$ has a short description? For example, suppose the entries $s_x$ of $\bs$, for $x \in [N]$, are given by an efficiently computable function $f(x)=s_x$, whose description size is much smaller than the size of $\bs$. In this case, we can hope to ``split'' $\bs$ into two {\em compressed} shares, describing functions $f_0$ and $f_1$, for which $s_x = f(x) = f_0(x)+f_1(x)$ for each $x \in [N]$, and where the description of each $f_i$ on its own computationally hides $f$ from within the class.  This is precisely the notion of {\em function secret sharing} (FSS)~\cite{EC:BoyGilIsh15}.

\paragraph{Distributed point functions (DPF).} It turns out that many cryptographic applications of FSS apply to functions with a small number of nonzero outputs. Such an FSS scheme provides a compressed additive secret-sharing of {\em sparse} vectors~$\bs$.
In the extreme version, $\bs$ contains at most one nonzero entry. Providing efficient, compressed secret shares of such a ``one-hot vector'' is captured by the notion of a {\em distributed point function} (DPF)~\cite{EC:GilboaIshai14,CCS:BoyGilIsh16}, which will serve as the core focus of this article.   

\subsection{DPF Definition}

It will be useful to consider the case where $N=2^n$ and view a maximally-sparse vector $\bs\in\G^N$ as the truth-table of a point function $f_{\alpha,\beta}:\{0,1\}^n\to\G$ defined by
$$
f_{\alpha, \beta}(x) = 
\begin{cases} 
\beta & \text{if } x = \alpha \\
0 & \text{if } x \neq \alpha.
\end{cases}
$$

The above functional view allows us to consider efficient random access to an output of a point function with an exponential input domain size $N$. This is enabled by allowing a succinct representation of the additive shares of $f_{\alpha,\beta}$ by a pair of keys $(k_0,k_1)$, generated by an efficient randomized key generation algorithm $\Gen(1^\lambda,\alpha,\beta)$, where $\lambda$ is a security parameter (typically $\lambda=128$ in practice). Random access to the additive shares of $f_{\alpha, \beta}(x)$ is enabled by another efficient algorithm $\Eval(i,k_i,x)$, where $i\in\{0,1\}$ is the key identity. This DPF syntax naturally gives rise to the following correctness and security requirements:
\begin{itemize}
    \item \textbf{Correctness:} For any input $x$, the reconstruction of the shares yields the function value:
    $$ \mathsf{Eval}(0, k_0, x) + \mathsf{Eval}(1, k_1, x) = f_{\alpha, \beta}(x) $$
    \item \textbf{Security:} A single key $k_i$ reveals essentially nothing about the secret index $\alpha$ or the output value $\beta$, in the sense that for any two pairs $(\alpha,\beta)$ and $(\alpha',\beta')$, the corresponding keys $k_i$ and $k'_i$ are computationally indistinguishable.
\end{itemize}
It is convenient to view the keys $(k_0,k_1)$ as succinctly representing {\em additive function shares} $(f_0,f_1)$, such that $f_0+f_1=f_{\alpha,\beta}$ and each $f_i$ computationally hides $f_{\alpha,\beta}$.

\subsection{Known Bounds on DPF Key Size} 

A trivial DPF construction that provides perfect (rather than computational) security is to additively share the truth-table of $f_{\alpha, \beta}$. This yields a DPF in which each key contains $N=2^n$ group elements. It turns out that significantly improving over this bound is impossible in the information-theoretic setting,  and moreover implies a PRG in the computational security setting~\cite{EC:GilboaIshai14}. Perhaps more surprisingly, a PRG is also sufficient for exponential compression. Concretely, using a PRG with a $\lambda$-bit seed, the best known DPF construction has keys of size $\approx n\cdot\lambda+\log|\G|$ bits~\cite{CCS:BoyGilIsh16}.

\paragraph{Construction at a glance.}
The state-of-the-art PRG-based DPF construction~\cite{CCS:BoyGilIsh16} uses additive secret sharing over $\mathbb{F}_2$, where a bit-string $s$ is split into two random strings whose XOR is $s$. It employs two simple ideas.  First, additive secret sharing is linearly homomorphic. This implies that parties can locally add public
corrections to shared values conditioned on shared control bits, viewing them as $\mathbb{F}_2$-affine functions of the shared bits.  Second, in
the two-party case, equal PRG seeds expand to equal strings, which cancel under
XOR, while independent seeds expand to pseudorandom-looking shares.  The
tree-based DPF combines these ideas level by level: all paths away from the
secret point are forced to cancel, while the unique path to the secret point
keeps independent pseudorandom seeds until a final correction installs the
payload $\beta$.  See Section~\ref{sec:construct} for more details.

\subsection{Application: PIR by Keywords} 
\label{sec:keywords}

To give a simple example for the usefulness of DPFs, consider the following keyword-search variant of 2-server {\em private information retrieval} (PIR)~\cite{CGKS95,KO97,ChorG97}. There are two servers $S_0$ and $S_1$ who each hold a database $D$ of keywords $x_j\in\{0,1\}^n$ (e.g., hashes of breached passwords).
A client, holding a search word $\alpha\in\{0,1\}^n$, wants to learn whether $\alpha\in D$
without revealing $\alpha$ to any server and while minimizing communication cost. A DPF gives rise to the following simple solution: The client views its search query as a point function $f_{\alpha,\beta}:\{0,1\}^n\to\Z_2$, where $\alpha$ is the secret search word and $\beta=1$. The client then uses $\Gen$ to split $f_{\alpha,\beta}$ into additive shares $f_{\alpha,\beta}=f_0+f_1$, sending the corresponding keys to the two servers. Each server $i$ uses $\Eval$ to compute $\sum_{x_j\in D} f_i(x_j)$, returning the one-bit sum $y_i\in\Z_2$ to the client. The client concludes that $\alpha\in D$ if $y_0+y_1=1$, and that $\alpha\not\in D$ otherwise. 

Note that the above application crucially relies on the {\em additive} representation of the output shares produced by $\Eval$ to aggregate the shared outputs over many inputs $x_i$. This application also demonstrates the importance of the random access provided by $\Eval$. When $n$ is large, as in the breached passwords example, scaling linearly with the domain size $N=2^n$ would be infeasible. We will later see additional applications of DPF that benefit from the above features.


\subsection{Multi-Party DPF} 

DPF can also be generalized from two parties to $m\ge 3$ parties.  The
efficiency landscape depends strongly on the corruption threshold $t$.

In the \emph{full-threshold} setting, $t=m-1$, security requires that any
strict subset of the $m$ keys hide the point function.  This is the natural
analogue of two-party DPF security, but it is much less well understood:
the best known PRG/OWF-based constructions have square-root dependence on
the domain size~\cite{EC:BoyGilIsh15,goel2025multiparty}, and beating this barrier is open even for three parties.

A different and more efficient regime is obtained when there is a gap between
the number of parties and the corruption threshold, namely $t<m-1$.  In this
setting one can sometimes obtain the stronger notion of \emph{multiplicative
DPF}, where the output shares support local multiplication of several DPF
outputs~\cite{bunn2022cnf,C:ABG+24}.  More precisely, an $(m,t)$-multiplicative DPF supports local
evaluation of polynomials of degree $\lfloor (m-1)/t\rfloor$ in the shared
DPF outputs; for $t=1$ this gives degree $m-1$.

Finally, in honest-majority-like regimes there are nontrivial
information-theoretic multi-party DPF constructions, discussed in
Section~\ref{sec:itdpf}.  Section~\ref{sec:multi-partyDPF} gives a more detailed comparison of the
full-threshold, gapped-threshold, and multiplicative multi-party settings.

\subsection{Function Secret Sharing} 

As mentioned above, a powerful generalization of DPF considers additively sharing functions $f$ from an arbitrary function class $\cal F$, rather than just point functions. This is referred to as a {\em function secret sharing} (FSS) scheme for $\cal F$~\cite{EC:BoyGilIsh15}. 

FSS schemes can naturally extend the functionality of most DPF applications. For example, in the above keyword search example, if we interpret each keyword as an $n$-bit integer, then FSS for the class of intervals (namely, functions that return 1 on an interval $[a,b]$ and 0 elsewhere) can be used to privately search for a database entry containing a number in a secret interval. While for this simple class $\cal F$ an efficient PRG-based FSS scheme is known~\cite{CCS:BoyGilIsh16,EC:BCG+21}, extending this to more general classes is a challenging open problem. Such extensions are known under different assumptions that imply public-key cryptography~\cite{EC:BoyGilIsh15,C:BoyGilIsh16,C:DHRW16,EC:BKS19,EC:OrlSchYak21,C:RoySin22,C:BCGIKS19,AbramDOS22}, though the resulting FSS schemes are typically much worse than PRG-based FSS in terms of concrete efficiency. We will discuss the general landscape of known FSS constructions in Section~\ref{sec:FSSconstruct}.

\subsection{Homomorphic Secret Sharing} 
\label{sec:HSS}

When considering FSS schemes for rich function classes, it is often useful to reverse the role of the function $f$ and the input $x$. This is captured by the notion of {\em homomorphic secret sharing} (HSS)~\cite{Ben86,C:BoyGilIsh16,BGILT18}. In an HSS scheme, a secret input $x$ is split between two or more parties, such that additive shares of $f(x)$ can be locally computed from the shares of $x$ for any (publicly known) function $f\in\cal F$. This dual form of FSS is more useful for some applications, and can be viewed as a natural distributed analogue of fully homomorphic encryption.

\subsection{Pseudorandom Correlation Generators} 
\label{sec:PCG-intro}

Finally, an increasingly popular application of DPFs is for compressing useful sources of correlated randomness. Consider first the simple case of encryption. Using a trusted dealer that distributes to a sender $P_0$ and a receiver $P_1$ the same secret random string $R\in\{0,1\}^N$, we enable $P_0$ to securely communicate a message $M\in\{0,1\}^N$ to $P_1$ by sending $M\oplus R$ over a public channel. Settling for computational security, this simple two-party correlation can be compressed by having the dealer send to both parties a short PRG seed $s$, which can be expanded to $R$. 

But what if we consider more general two-party correlations, such as ones that can be used for secure two-party computation? For example, we can consider an ideal target correlation $(R_0,R_1)$ consisting of $N$ independent instances of random {\em  oblivious transfer} (OT), where for each instance $R_0$ includes a pair of random bits and $R_1$ includes a random selection of one of these bits. Can we securely compress such a two-party correlation by having the dealer generate a pair of {\em short} correlated seeds $(s_0,s_1)$ that can be locally expanded into outputs that are indistinguishable from the target correlation? Note that here security means that knowing one seed $s_i$ does not reveal more information than is implied by the target correlation. In the random OT example, this means that $s_0$ should hide the selections of $P_1$, and $s_1$ should hide the bits of $R_0$ that were not selected. 

This type of compression is captured by the notion of a {\em pseudorandom correlation generator} (PCG)~\cite{CCS:BoyleCGI18,C:BCGIKS19}. It turns out that DPFs can be used to build practical PCGs for useful correlations, including the above random OT correlation, based on different flavors of the \emph{learning parity with noise} (LPN) assumption. The high level idea is to use DPF for compressing a variant of the correlation that involves {\em sparse} random vectors, and then use a public linear mapping to convert the sparse correlation into a pseudorandom correlation. 

See Section~\ref{sec:PCG} for further discussion of the current landscape of PCG constructions.

\subsection{Organization}
This survey provides an introduction to distributed point functions and related primitives, as well as an overview of the state of the art within this research domain. Section~\ref{sec-defs} establishes the formal definitions and security models. Section~\ref{sec:constructions} discusses constructions of DPFs and their variants. Section~\ref{sec:apps} discusses different applications of DPFs in cryptography and beyond. Finally, Section~\ref{sec:openQ} presents a collection of open problems in the area.

Table~\ref{tab:fss-techniques} gives a high-level map of the FSS landscape, including the general FSS abstraction, common concrete function classes that instantiate it, and related primitives, such as HSS and PCGs.

\definecolor{GeneralFSSBg}{RGB}{250,250,252}
\definecolor{GeneralFSSHeader}{RGB}{238,240,245}
\definecolor{FSSExampleBlue}{RGB}{242,247,255}
\definecolor{HeaderBlue}{RGB}{225,235,250}
\definecolor{RelatedGray}{RGB}{247,247,247}
\definecolor{HeaderGray}{RGB}{232,232,232}

\newcommand{\GroupHeader}[1]{\multicolumn{5}{@{}l}{\textbf{#1}}}
\newcommand{\RowTitle}[2]{\textbf{#1} {\normalfont(\S#2)}}


\begin{table*}[t]
\centering
\caption{
High-level map of the FSS landscape. Efficient DPF, DCF, DMPF, Multiplicative DPF, and verifiable/extractable DPF exist from any pseudorandom generator (PRG). General FSS, offset-function FSS, HSS, PCG/PCFs exist from a range of assumptions, depending on the complexity of functions supported. Many useful efficient PCGs rely on the Learning Parity with Noise (LPN) assumption and variants.
}
\label{tab:fss-techniques}
\renewcommand{\arraystretch}{1.18}
\setlength{\tabcolsep}{4pt}
\scriptsize
\begin{tabularx}{\textwidth}{
  >{\raggedright\arraybackslash}p{0.15\textwidth}
  >{\raggedright\arraybackslash}p{0.21\textwidth}
  >{\raggedright\arraybackslash}p{0.16\textwidth}
  >{\raggedright\arraybackslash}p{0.23\textwidth}
  >{\raggedright\arraybackslash}X
}
\toprule
\textbf{Primitive / notion} &
\textbf{Secret object} &
\textbf{Reconstruction (input $x$)} &
\textbf{Core technique} &
\textbf{Representative uses} \\
\midrule

\rowcolor{GeneralFSSHeader}
\GroupHeader{General FSS notion} \\
\addlinespace[0.15em]

\rowcolor{GeneralFSSBg}
\RowTitle{FSS for function class $\mathcal{F}$}{\ref{sec:def-FSS}}  &
A secret function $f \in \mathcal{F}$ &
Additive shares of $f(x)$ for each public input $x$ &
Key generation splits $f$ into compact function shares $f_0,f_1$
such that $f_0(x)+f_1(x)=f(x)$, while each key hides $f$ up to
allowed leakage &
Generic abstraction for private evaluation of secret-shared functions;
specialized efficient schemes exploit the structure of $\mathcal{F}$ \\

\addlinespace[0.45em]
\rowcolor{HeaderBlue}
\GroupHeader{Examples of FSS function classes and variants} \\
\addlinespace[0.15em]

\rowcolor{FSSExampleBlue}
\RowTitle{DPF}{\ref{sec:def:DPF}, \ref{sec:construct}} &
Point function $f_{\alpha,\beta}$, equivalently a weight-one vector $f_{\alpha,\beta}(\alpha)=\beta$ &
Additive shares of $f_{\alpha,\beta}(x)$ &
PRG tree with level-by-level correction words; off-path branches cancel &
PIR, private writes, histograms, equality tests \\

\rowcolor{FSSExampleBlue}
\RowTitle{DCF}{\ref{sec:extensions}, \ref{sec:DCF}} &
Comparison function $f^{<}_{\alpha,\beta}$, where
$f^{<}_{\alpha,\beta}(x)=\beta$ for $x<\alpha$ &
Additive shares of $f^{<}_{\alpha,\beta}(x)$ &
Prefix-DPF structure or optimized tree-based comparison construction &
Range queries, interval functions, comparisons in secure computation \\

\rowcolor{FSSExampleBlue}
\RowTitle{DMPF}{\ref{sec:extensions}, \ref{sec:DMPF}} &
Multi-point (sparse) function $f_{A,B}$, with up to $t$ nonzero points ($t$ is revealed) &
Additive shares of $f_{A,B}(x)$ &
Sum of DPFs; big-state DPF; probabilistic batch codes; OKVS encodings &
Pseudorandom correlation generators, unbalanced PSI, sparse-vector sharing \\

\rowcolor{FSSExampleBlue}
\RowTitle{Multiplicative DPF}{\ref{sec:extensions}, \ref{sec:multi-partyDPF}} & 
Point function, but where output shares support local multiplication (requiring gap between corruption threshold and number of servers) &
Multiplicative secret sharing, followed by additive reconstruction of products &
Tree-based DPF combined with CNF, replicated, or Shamir-style output sharing &
Conjunctive queries, low-depth products of DPF outputs, unit-vector correlations \\

\rowcolor{FSSExampleBlue}
\RowTitle{Offset-function FSS}{\ref{sec:preprocessing}} &
Offset gate function for public gate $g$ and secret masks
$r_{\mathsf{in}},r_{\mathsf{out}}$:
$g_{r_{\mathsf{in}},r_{\mathsf{out}}}(\hat{x})
 = g(\hat{x}-r_{\mathsf{in}})+r_{\mathsf{out}}$ &
Additive shares of the masked gate output &
FSS for structured offset classes induced by equality, comparison,
truncation, ReLU, splines, and related gates &
Secure computation with preprocessing, secure mixed-mode computation, private ML inference \\

\addlinespace[0.45em]
\rowcolor{HeaderGray}
\GroupHeader{Related concepts} \\
\addlinespace[0.15em]

\rowcolor{RelatedGray}
\RowTitle{HSS}{\ref{sec:HSS}, \ref{sec:FSSconstruct}} &
Secret input $x$ evaluated under a public function $f$ &
Additive or structured shares of $f(x)$ &
Dual view of FSS: the input is secret-shared and the function is public &
Secure computation, public-key-style FSS, richer function classes \\

\rowcolor{RelatedGray}
\RowTitle{PCG / PCF}{\ref{sec:PCG-intro}, \ref{sec:PCG}} &
Large correlated randomness expanded from short correlated seeds &
Local expansion to correlated outputs such as Oblivious Transfer (OT), or Beaver triples &
DPF/DMPF sharing of sparse errors followed by public linear maps from
LPN-style assumptions; PCFs give a function-like analogue &
Silent OT extension, Beaver triples, preprocessing for secure computation \\

\rowcolor{RelatedGray}
\RowTitle{Verifiable / extractable DPF}{\ref{sec:private-writing}} &
A well-formed point function, even against maliciously generated keys &
Additive shares plus a validity or extractability guarantee &
Arithmetic sketching, hash-based consistency checks, or random oracle-based
extractability of the DPF tree &
Malicious-client private aggregation, heavy hitters, anonymous messaging \\

\bottomrule
\end{tabularx}
\end{table*}

\section{Definitions}
\label{sec-defs}

We begin by giving a formal definition of two-party function secret sharing (FSS). We then zoom into the corresponding special case of two-party distributed point functions (DPF), which serves as the predominant focus of this survey. We then discuss useful extensions, including variants with more than two parties; programmable DPF, where one key can be generated before the secret point function is selected; and extractable DPF, providing tighter guarantees on the well-formedness of DPF keys.

\paragraph{Notation and algebraic conventions.}
We use $\lambda$ for the security parameter throughout, and write
probabilities as $\Pr[\cdot]$.  All output groups are Abelian and written
additively unless stated otherwise.  Thus $0$ denotes the identity element
and $-a$ denotes the additive inverse of $a$.  We write $+$ for the general
group operation and reserve $\oplus$ for bitwise XOR over groups such as
$\mathbb{Z}_2^\ell$.

A ring $R$ is a set equipped with addition and multiplication, such as
$\mathbb{Z}_q$.  An $R$-module is the analogue of a vector space in which
scalars come from $R$ rather than necessarily from a field.  For example,
$\mathbb{Z}_q^d$ is a module over $\mathbb{Z}_q$.  Thus an expression such as
$\sum_i \alpha_i g_i(x)$ means that coefficients $\alpha_i\in R$ multiply
public values $g_i(x)$ in an $R$-module, and the results are added using the
module addition.

\subsection{Function Secret Sharing}
\label{sec:def-FSS}

We follow the definition of function secret sharing (FSS) from~\cite{CCS:BoyGilIsh16}. Intuitively, a (2-party) FSS scheme is an efficient algorithm that splits a function $f\in\cF$ into two {\em additive} shares $f_0,f_1$, such that: (1) each $f_i$ hides $f$ beyond some allowable leakage information $\leak(f)$; and (2) for every input $x$, $f_0(x)+f_1(x)=f(x)$. 
The main challenge is to make the descriptions of $f_0$ and $f_1$ compact, while still allowing their efficient evaluation. 


\begin{definition}[FSS: Syntax] \label{def:FSS-syntax}
A ($2$-party) {\em function secret sharing (FSS) scheme} is a pair of  algorithms $(\gen,\eval)$ such that:
  \begin{itemize}
  
  \item $\gen(1^\lambda, f)$ is a PPT {\em key-generation algorithm} that given security parameter $1^\lambda$ and description of a function $f \in \zo^*$ outputs a pair of keys $(k_0,k_1)$. We assume that $f$ explicitly contains descriptions of input and output groups $\G_\gin,\G_\gout$ of $f: \G_\gin \to \G_\gout$.
  
  \item $\eval(i, k_i, x)$ is a polynomial-time {\em evaluation algorithm} that given party index $i\in\zo$, key $k_i$ (defining $f_i:\G_\gin\to\G_\gout$), and input $x \in \G_\gin$ outputs a group element $y_i \in \G_\gout$ (the value of $f_i(x)$).  
  \end{itemize}
\end{definition}

\begin{definition}[FSS: Correctness and Security] \label{def:FSS-semantics}
Let $\cF$ be a function family and $\leak$ be a function specifying the allowable leakage about $f \in \cF$. When $\leak$ is omitted, it is understood to output only a description of $\G_\gin$ and $\G_\gout$. 
We say that $(\gen,\eval)$ as in Definition~\ref{def:FSS-syntax} is an {\em FSS scheme for $\cF$} (with respect to leakage $\leak$) if it  satisfies the following requirements.
\begin{itemize}
 
  \item {\bf Correctness:} For all $f \in \cF$ describing $f:\G_\gin\to\G_\gout$, and every $x \in \G_\gin$, if $(k_0,k_1) \gets \gen(1^\lambda,f)$, then
  $$\Pr \left[\eval(0,k_0,x)+\eval(1,k_1,x)  = f(x) \right] = 1.$$
	
  \item {\bf Security:} For each $i\in\zo$, there is a PPT algorithm $\Sim_i$ (simulator), such that for every sequence $(f_\lambda)_{\lambda\in\N}$ of polynomial-size function descriptions from $\cF$,
  the outputs of the following experiments $\Real$ and $\Ideal$ are computationally indistinguishable:
    \begin{itemize}
    \item 
$\Real_\lambda$: $(k_0,k_1) \gets \gen(1^\lambda, f_\lambda)$; Output $k_i$.

    \item 
  $\Ideal_\lambda$: Output $\Sim_i(1^\lambda, \leak(f_\lambda))$.
    \end{itemize}
  \end{itemize}
\end{definition}

\paragraph{Simulation vs.\ indistinguishability security.} In contrast to the simulation-based security requirement above, the original FSS definition from~\cite{EC:BoyGilIsh15}
was formulated as a semantic-security notion of indistinguishability. Namely, security required that for each party $i \in \zo$, for any two sequences $(f_\lambda)_{\lambda\in\N},(f'_\lambda)_{\lambda\in\N}$ of polynomial-size function descriptions from $\cF$, then party $i$'s key share $k_i$ as generated from $(k_0,k_1) \gets \gen(1^\lambda,f_\lambda)$ is indistinguishable from its key as generated from $(k_0,k_1) \gets \gen(1^\lambda,f'_\lambda)$.


Interestingly, the two flavors of definition are equivalent for 
any function family $\cal F$ and leakage function $\Leak$ for which $\Leak$ can be efficiently inverted. More concretely, suppose given $\Leak(f)$ one can efficiently find $f' \in \cF$ such that $\Leak(f')=\Leak(f)$. Then the required simulator $\Sim_i(1^\lambda,z = \leak(f_\lambda))$ can be attained by choosing an arbitrary $f'_\lambda \in \cF$ for which $\leak(f'_\lambda) = z$, honestly generating keys $(k'_0,k'_1) \gets \gen(1^\lambda,f'_\lambda)$, and outputting the corresponding key $k'_i$ as its simulation.  
If the FSS scheme has indistinguishability security, then this simulated key $k'_i$ is indistinguishable from the true key $k_i$ generated from $f_\lambda$, as desired. In the other direction, the simulation requirement automatically implies indistinguishability of party $i$'s key from any choice of secret function $f$ with the same value of $\leak(f)$.

Such an inversion algorithm exists for all instances of $\cal F$ and $\Leak$ considered in this survey---including distributed point functions and extensions---and essentially all from the literature.  

\paragraph{Post-quantum security.} While in this survey we focus on classical security, one can naturally consider FSS with post-quantum security. Almost all of the constructions we discuss (with the exception of group-based FSS schemes discussed in
Section~\ref{sec:FSSconstruct}) are plausibly post-quantum secure.

\subsection{Distributed Point Function}
\label{sec:def:DPF}

We next give a formal definition of distributed point functions (DPF), as formulated in~\cite{CCS:BoyGilIsh16}. Roughly speaking, DPF is an instance of function secret sharing as in the previous section, for the class $\cF$ of {\em point functions}, and with $\leak(f)$ revealing the input and output space of the corresponding point function $f$.

\begin{definition}[Point Functions] 
A {\em point function} $f_{\alpha,\beta}$, for $\alpha\in\zo^{n}$ and $\beta\in\bbG$, is defined to be the function $f:\zo^n\to\bbG$ such that $f(\alpha)=\beta$ and $f(x)=0$ for $x\neq \alpha$. Here, 0 represents the identity of the abelian group~$\G$.
\end{definition}

\newcommand{\fab}{f_{\alpha,\beta}}

\begin{definition}[DPF]
A (2-party) distributed point function (DPF) scheme is a pair of efficient algorithms $(\gen,\eval)$ with the following syntax: 
  \begin{itemize} 
      \item $\gen(1^\lambda, \fab)$ is a {\em key generation algorithm} that given security parameter $1^\lambda$  and description of a point function $\fab: \zo^n \to \G$ outputs a pair of keys $(k_0,k_1)$. 
      We assume that $\fab$ explicitly contains the input length $n$ and the output group $\G$. 

      \item $\eval(i, k_i, x)$ is an {\em evaluation algorithm} that given party index $i\in\zo$, key $k_i$, and $x \in \zo^n$ in the domain of $\fab$, outputs a group element $y_i \in \G$.  
  \end{itemize}
The algorithms $(\gen,\eval)$ should satisfy the following two properties:
  \begin{itemize}
  \item {\bf Correctness}:
  For every $\fab: \zo^n \to \G$ and input $x \in \zo^n$, if $(k_0,k_1) \gets \Gen(1^\lambda,\fab)$ then the corresponding evaluated output shares satisfy:
	\[\Pr \left[  \Eval(0,k_0,x) + \Eval(1,k_1,x)  = f(x) \right] = 1.\]

  \item {\bf Security:} 
  For every $i^* \in \zo$, there exists a PPT algorithm $\Sim$ (simulator), such that for every 
  polynomial-size point function $\fab$,
  the outputs of the following experiments $\Real$ and $\Ideal$ are computationally indistinguishable:
  \begin{itemize}
  \item 
$\Real(1^\lambda)$: $(k_0,k_1) \gets \Gen(1^\lambda, \fab)$;  Output $k_{i^*}$.

\item 
  $\Ideal(1^\lambda)$: Output $\Sim(1^\lambda, (1^n,\G))$.
  \end{itemize}
  
 \end{itemize}
\end{definition}


\paragraph{Additional syntax.} Besides the usual $\gen$ and $\eval$ algorithms, DPF schemes are sometimes equipped with the following additional syntax, for use in applications:

\begin{itemize} 
    \item $\distgen(1^\secp, \fab)$: A \emph{distributed key generation protocol}, which securely realizes the functionality that takes from $P_0$ and $P_1$ shares of $\alpha,\beta$ as inputs, evaluates $\Gen(1^\lambda,f_{\alpha,\beta})$, and outputs key $k_i$ to party $P_i$. 
    We will give examples of such protocols in Section~\ref{sec:DistribGen}. 
    
    \item $\fulleval(i, k_i)$: A \emph{full-domain evaluation algorithm} that given a party index $i \in \zo$ and key $k_i$, outputs a vector of group elements $(y_i^x)_{x \in \zo^n}$ s.t. $y_0^x + y_1^x = f(x)$ for all $x \in \zo^n$. It can be seen as an enhanced form of the $\eval$ algorithm which, instead of evaluating at a chosen input $x \in \zo^n$, evaluates at {\em all} inputs in $\zo^n$. This algorithm is feasible only when the DPF has polynomial-size input domain, and is beneficial whenever there is a way to amortize the costs of multiple $\Eval$ instances.
\end{itemize}

In the usual client-generated PDF setting, $\Gen$ is run locally by a client who
knows $(\alpha,\beta)$.  By contrast, $\distgen$ is needed when $\alpha$ and/or
$\beta$ are themselves secret-shared among parties.  This is the case, for example, when jointly generating seeds of a pseudorandom correlation generator (PCG), or performing secure computation of RAM programs~\cite{doerner2017scaling}.   In such case, the parties must run an interactive secure
protocol for the relevant $\Gen$ functionality.  A generic secure computation
of the $\Gen$ circuit is possible, but is usually unattractive because it is
non-black-box in the underlying cryptographic tools (e.g., pseudorandom generators).  Section~\ref{sec:DistribGen} discusses the standard
black-box alternative based on local PRG calls and oblivious transfer.

Full-domain evaluation is useful for applications such as private information retrieval, where the domain size $N=2^n$ corresponds to the (polynomial-size) database size, and secret shares of the full expanded unit vector are required for functionality.  For the tree-based DPF construction, $\fulleval$ expands the 
full construction tree once and costs $O(N)$ PRG calls plus $O(N)$ output writes;
it can be implemented with $O(N)$ memory if all leaves are materialized, or
with small working memory if leaves are streamed.  This is in contrast to $N$ individual calls to $\eval$, which would require larger cost $O(N \log N)$.



\subsection{Extensions Beyond 2-Party DPF}
\label{sec:extensions}

The above definition of 2-party DPF can be generalized in several orthogonal ways. This includes, in particular:
\begin{itemize}
    
    \item {\em To $m>2$ parties:}  FSS, including DPF, can be generalized to $m \ge 2$ parties with $t < m$ corruptions, where security requires that any subset of $t$ keys reveal no information beyond the specified leakage $\leak(f)$.
    
    \item {\em Non-additive reconstruction:} 
    The additive reconstruction property $\Eval(0,k_0,x)+\Eval(1,k_1,x) = f(x)$ of FSS (more generally, its linear structure) is what enables many of its applications. 
    Linearity of reconstruction provides convenient share compressibility;  for example, $\sum_{x \in S} \Eval(0,k_0,x)$ and $\sum_{x \in S} \Eval(1,k_1,x)$ automatically form shares of $\sum_{x\in S} f(x)$.
    Further, output shares must themselves be elements of the function output space, immediately guaranteeing share succinctness. 

    Nontrivial notions of FSS can be considered with relaxed requirements on reconstruction, which may suffice for certain applications.
    For example, a natural requirement is that the output length of Eval is short, depending only on the output length of $f$ and not on the input length of~$f$.


\end{itemize}


We refer the reader to~\cite{EC:BoyGilIsh15,BGILT18} for a formal treatment of these and further extensions. In this survey, we consider only additive reconstruction.  We will focus predominantly on the two-party setting, addressing the case with more parties only briefly (see Section~\ref{sec:multi-partyDPF}). And, we will mostly consider the case of DPF as well as the following highly related variants.

\paragraph{Distributed {\em Comparison} Function (DCF).}
A DCF is an FSS for the class of {\em comparison} functions $f_{\alpha, \beta}^{<}: \zo^n \to \G$, where $\alpha \in \zo^n$, $\beta \in \G$, and $f_{\alpha, \beta}^{<}(x)= \beta$ for $x < \alpha$ and $0 \in \G$ for $x\ge \alpha$ (where ordering on $\zo^n$ is given by integer interpretation, i.e., the bit-string $0^n$, corresponding to the integer $0$, is the smallest element and the bit-string $1^n$, corresponding to the integer $2^n - 1$, is the largest element). 

We discuss constructions of DCF in Section~\ref{sec:constructions}.

\paragraph{Distributed Multi-Point Functions (DMPF).}
Many applications of DPF (e.g., for constructions of pseudorandom correlation generators as in Section~\ref{sec:PCG}) require compressed shares of a sparse weight-$t$ vector. This corresponds to secret shares of a {\em $t$-point function} in the place of a (weight-1) point function.

 Formally, a \emph{$t$-point function} $f_{A,B}:\zo^n\rightarrow \bbG$ for $A=\{\alpha_1,\cdots\alpha_t\}\subset \zo^n$ listed in ascending order and $B=(\beta_1,\cdots,\beta_t)\in \bbG^t$ evaluates to $\beta_i$ on input $\alpha_i$ for $1\le i\le t$ and to $0$ on all other inputs. 
 We denote by \emph{multi-point functions} the collection of all $t$-point functions for all $t$. A Distributed {\em Multi}-Point Function (DMPF) is an FSS scheme for this function class with leakage function ${\sf Leak}(f_{A,B}) = (1^n,\bbG,t)$. In particular, the FSS keys are allowed to reveal a bound on the weight $t$, and the complexity of the scheme is allowed to grow with~$t$.


    

\paragraph{Multiplicative DPF.}
A natural variant of DPF, considered in~\cite{bunn2022cnf,C:ABG+24}, is $(m,t)$-multiplicative DPF.  Here, the output group $\G$ is the additive group of a commutative ring $R$, and the outputs of $\eval$ on input $x$ are an $(m,t)$-multiplicative secret sharing of the desired point function evaluation value $f_{\alpha,\beta}(x)$.
 Security requires that any subset of $t$ FSS keys
reveals no information about $f_{\alpha,\beta}$ beyond $\Leak(f_{\alpha,\beta})$.
The multiplicative output sharing allows the parties to compute any polynomial
of degree at most $\lfloor (m-1)/t\rfloor$ in the shared DPF output by local
computation on their shares.  In particular, when $t=1$, this degree is $m-1$.

\section{Constructions}
\label{sec:constructions}

In this section, we address known constructions of FSS schemes, focusing on DPF.  We begin with a detailed treatment of the best current construction of two-party DPF (Section~\ref{sec:construct}) and related extensions: multi-party DPF (Section~\ref{sec:multi-partyDPF}), distributed {\em comparison} functions (DCF; Section~\ref{sec:DCF}), distributed {\em multi-point} functions (DMPF; Section~\ref{sec:DMPF}), and information-theoretic DPF (Section~\ref{sec:itdpf}).
We then conclude with a general overview of known FSS constructions for different function classes (Section~\ref{sec:FSSconstruct}).

\subsection{Constructing DPF}
\label{sec:construct}

In this section, we describe the state-of-the-art constructions of distributed point function (DPF) and distributed comparison function (DCF) from any pseudorandom generator (PRG). 
We refer the reader also to~\cite{BIU-talk} for a video lecture covering these constructions as an additional resource.

\subsubsection{DPF Warmup: Building Blocks and Square-Root DPF}
\label{sec:warmup}

To illustrate the ideas behind the two-party DPF construction, consider the following intuition. All constructions are based on two simple building blocks. The first is {\em additive secret sharing}, which takes a secret over an Abelian group and splits it into two random group elements that add up to the secret. In our case, the Abelian group will typically be of the form $\mathbb{Z}_2^\ell$, namely the set of $\ell$-bit strings with the XOR operation. We view the secret as being split between two parties and denote the pair of shares of a secret $s\in\{0,1\}^\ell$ by $[s]=(s_0,s_1)$. The second building block is a PRG $G$ that stretches a $\lambda$-bit random seed to a pseudorandom string of polynomial length $\ell(\lambda)\gg \lambda$. 

Now, let's consider the following basic question: Which kinds of homomorphic operations can we perform on an additively shared secret $s\in\{0,1\}^\ell$? More concretely, for which functions $f$ can the parties locally convert $[s]$ into $[f(s)]$ without knowing $s$?  We have two simple kinds of homomorphisms:
\begin{enumerate}
\item
Additive secret sharing is {\em linearly homomorphic}: If $[s]=(s_0,s_1)$ are additive shares of $s$ then $A\cdot[s]=(A\cdot s_0,A\cdot s_1)$ are additive shares of $As$. A useful corollary is the following  {\em conditional correction} gadget: given additive shares $[s],[t]$ of a string $s\in\{0,1\}^\ell$ and a ``control bit'' $t\in\{0,1\}$, along with a public correction word $CW\in\{0,1\}^\ell$, the parties can locally compute $[s']$ for $s'=s+t\cdot CW$. We view this as a conditional correction of the secret $s$ by $CW$ conditioned on $t=1$.  
\item
In the 2-party case, additive secret sharing satisfies the following weak homomorphism: If $G:\{0,1\}^\lambda\to\{0,1\}^\ell$ is a PRG, then $G([s])=(G(s_0),G(s_1))$ extends shares of the 0-string $s=0^\lambda$ into shares of a longer 0-string $s'=0^\ell$, and shares of a random string $s\in\{0,1\}^\lambda$ into shares of a longer (pseudo-)random string $s'$, where $s'$ is pseudo-random even given one share of $s$. 
\end{enumerate}

Jumping ahead, the fact that the second homomorphism only applies to the 2-party case explains why the best known DPF constructions do not extend beyond two parties.

\begin{figure*}[t]
\centering
\begin{tikzpicture}[
  seed/.style={draw, rounded corners=2pt, minimum width=1.3cm, minimum height=0.5cm, font=\small},
  same/.style={seed, fill=blue!10},
  diff0/.style={seed, fill=red!15},
  diff1/.style={seed, fill=orange!15},
  mcell/.style={draw, minimum size=0.55cm, font=\small, inner sep=0pt},
  >=stealth,
]

\def\ya{2.4}   
\def\yva{1.6}  
\def\yb{0.8}   
\def\yvb{0.0}  
\def\yc{-0.8}  


\node[font=\small\bfseries] at (0, 3.4) {$k_0$};
\node[font=\small\bfseries] at (2.8, 3.4) {$k_1$};
\node[font=\small\bfseries] at (4.4, 3.4) {$[t^j]$};

\node[font=\small, anchor=east] at (-1.2, \ya) {Row $1$:};
\node[font=\small, anchor=east, red!70!black] at (-1.2, \yb) {Row $i$:};
\node[font=\small, anchor=east] at (-1.2, \yc) {Row $\ell$:};

\node[same] at (0, \ya) {$s^1$};
\node[diff0] at (0, \yb) {$s^i_0$};
\node[same] at (0, \yc) {$s^\ell$};

\node[same] at (2.8, \ya) {$s^1$};
\node[diff1] at (2.8, \yb) {$s^i_1$};
\node[same] at (2.8, \yc) {$s^\ell$};

\node at (1.4, \ya) {$=$};
\node[red!70!black, font=\bfseries] at (1.4, \yb) {$\neq$};
\node at (1.4, \yc) {$=$};

\node[font=\small] at (4.4, \ya) {$[0]$};
\node[font=\small, red!70!black] at (4.4, \yb) {$[1]$};
\node[font=\small] at (4.4, \yc) {$[0]$};

\foreach \x in {0, 1.4, 2.8, 4.4} {
  \node at (\x, \yva) {$\vdots$};
  \node at (\x, \yvb) {$\vdots$};
}

\node[draw, fill=violet!8, rounded corners=3pt, font=\small] at (1.4, -1.8) {$CW = G(s^i_0) \oplus G(s^i_1) \oplus (0,\dots,\beta,\dots,0)$};

\draw[->, line width=1.5pt] (5.6, 0.8) -- node[above, font=\small, align=center] {PRG expand} node[below, font=\small, align=center] {+ correct} (7.2, 0.8);

\def\mx{9.5}
\def\cs{0.65}

\node[font=\small\bfseries] at (\mx+2*\cs, 3.4) {$M = M_0 \oplus M_1$ ~~$(\ell \times \ell)$};

\node[font=\scriptsize] at (\mx, 2.85) {col $1$};
\node[font=\scriptsize] at (\mx+\cs, 2.85) {$\cdots$};
\node[font=\scriptsize, green!40!black] at (\mx+2*\cs, 2.85) {col $j^*$};
\node[font=\scriptsize] at (\mx+3*\cs, 2.85) {$\cdots$};
\node[font=\scriptsize] at (\mx+4*\cs, 2.85) {col $\ell$};

\node[mcell, fill=gray!5] at (\mx, \ya) {$0$};
\node at (\mx+\cs, \ya) {$\cdots$};
\node[mcell, fill=gray!5] at (\mx+2*\cs, \ya) {$0$};
\node at (\mx+3*\cs, \ya) {$\cdots$};
\node[mcell, fill=gray!5] at (\mx+4*\cs, \ya) {$0$};

\foreach \c in {0, 2, 4} {
  \node at (\mx+\c*\cs, \yva) {$\vdots$};
}

\node[mcell, fill=gray!5] at (\mx, \yb) {$0$};
\node at (\mx+\cs, \yb) {$\cdots$};
\node[mcell, fill=green!25, font=\small\bfseries] at (\mx+2*\cs, \yb) {$\beta$};
\node at (\mx+3*\cs, \yb) {$\cdots$};
\node[mcell, fill=gray!5] at (\mx+4*\cs, \yb) {$0$};

\draw[red!60!black, thick, rounded corners=2pt]
  (\mx-0.35, \yb-0.35) rectangle (\mx+4*\cs+0.35, \yb+0.35);

\foreach \c in {0, 2, 4} {
  \node at (\mx+\c*\cs, \yvb) {$\vdots$};
}

\node[mcell, fill=gray!5] at (\mx, \yc) {$0$};
\node at (\mx+\cs, \yc) {$\cdots$};
\node[mcell, fill=gray!5] at (\mx+2*\cs, \yc) {$0$};
\node at (\mx+3*\cs, \yc) {$\cdots$};
\node[mcell, fill=gray!5] at (\mx+4*\cs, \yc) {$0$};

\node[font=\scriptsize, blue!60!black, anchor=west, align=left] at (\mx+4*\cs+0.6, \ya) {same seeds\\$\Rightarrow$ zero row};
\node[font=\scriptsize, red!60!black, anchor=west, align=left] at (\mx+4*\cs+0.6, \yb) {indep.\ seeds\\+ $CW$ correction};
\node[font=\scriptsize, blue!60!black, anchor=west, align=left] at (\mx+4*\cs+0.6, \yc) {same seeds\\$\Rightarrow$ zero row};

\end{tikzpicture}
\caption{Square-root DPF construction for input domain $[N]$ with $N = \ell^2$. \textbf{Left:} The two DPF keys $(k_0, k_1)$. Each key contains $\ell$ PRG seeds (one per matrix row) and secret-shared control bits $[t^j]$. For rows $j \neq i$, both keys share the same seed $s^j$; for the special row $i$ (containing target $\alpha$), the keys hold independent seeds $s^i_0 \neq s^i_1$. A public correction word $CW$ is included in both keys. \textbf{Right:} The reconstructed $\ell \times \ell$ matrix $M = M_0 \oplus M_1$. Shared seeds yield identical PRG outputs, so their XOR produces zero rows. Row $i$ expands to a pseudorandom row, corrected via $CW$ conditioned on $[t^i]=[1]$ to place $\beta$ at column $j^*$ (the position of $\alpha$ within row $i$). Key size: $(\lambda+1)\cdot\ell + \ell = O(\sqrt{N})$.}
\label{fig:sqrt-dpf}
\end{figure*}
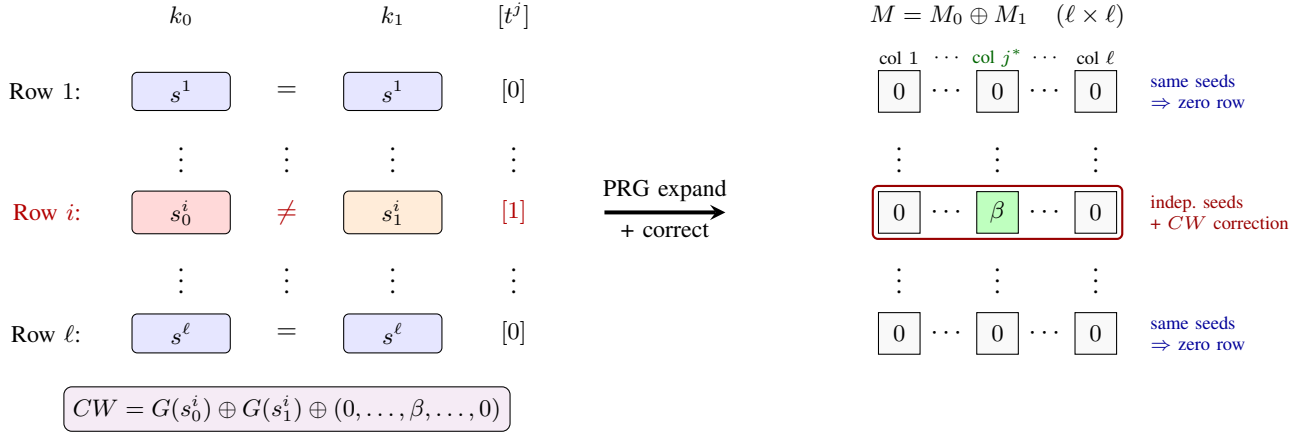

Armed with the above two types of homomorphism, we are ready to describe a simple ``square-root DPF'' construction, in which the key size scales linearly with the square-root of the input domain size $N=2^n$. This DPF is implicit in the 2-server PIR scheme of Chor and Gilboa~\cite{ChorG97}.
The construction is shown in Fig~\ref{fig:sqrt-dpf}.

Consider a DPF with input domain $[N]$, where $N=\ell^2$, and output group $\Z_2$. We can view the two keys as defining additive shares of a vector $u\in\Z_2^N$ such that $u_\alpha=\beta$ and $u_{\alpha'}=0$ for all $\alpha'\neq \alpha$. Folding $u$ into an $\ell\times \ell$ matrix $M$, coordinate $\alpha$ in $u$ belongs to some row $i\in[\ell]$ of $M$; all other rows of $M$ contain only zeros. As a first step, we generate compressed additive shares of a related matrix $M'$, which is similar to $M$ except that row $i$ is pseudorandom. This is done by choosing an independent PRG seed $s^j\in\{0,1\}^\lambda$ for each row $j\neq i$ of $M'$, and two independent seeds $s^i_0,s^i_1$ for row $i$. Now, letting each DPF key $k_b$ include the $\ell$ seeds $s^j_b$, where for $j\neq i$ we have $s^j_b=s^j$, we can apply the PRG-based homomorphism to each row, locally expanding the keys $(k_0,k_1)$ into additive shares $[M']$ as above. 

As the second and final step, we need to correct row $i$ of $M'$ to match row $i$ of $M$. To this end, we exploit the linear homomorphism of the additive sharing of $M'$ and apply the above correction gadget: for each row of $M'$ we add to the keys a secret-shared control bit, which is $1$ for row $i$ and $0$ for all other rows. Finally, we add a public correction word $CW\in\{0,1\}^\ell$ to correct row $i$ of $M'$ to match row $i$ in $M$. All in all, each augmented key includes $(\lambda+1)\cdot \ell$ secret bits, along with an $\ell$-bit public correction. Since $\ell=\sqrt{N}$, the key size scales linearly with $\sqrt{N}$.

Looking at the structure of the two keys, they are identical except in the length-$\lambda$ part that corresponds to row $i$. Thus, we can view the non-public part of these keys as an additively shared point function over $\G=\Z_2^{\lambda}$. This suggests the possibility of further compressing the non-public part via the use of recursion. In the following section, we describe a tree-based approach which is loosely based on this idea.

\subsubsection{Tree-Based DPF}
\label{sec:tree}

In this section we describe the state-of-the-art PRG-based DPF construction from~\cite{CCS:BoyGilIsh16}. Unlike the previous square-root construction, here the key size scales logarithmically with the domain size $N$, namely linearly in the input length $n$. The parameters of this construction are captured by the following theorem.

\begin{theorem}[Tree-based two-party DPF]   \label{thm:optimized-DPF}
Suppose $G: \{0,1\}^\lambda \to \{0,1\}^{2(\lambda+1)}$ is a pseudorandom generator.
Then the scheme $(\Gen^\pt,\Eval^\pt)$ from Figure~\ref{fig:DPFx4} is a 
DPF for the family of point functions $f_{\alpha,\beta} : \{0,1\}^n \to \bbG$
 with key size 
 $n\cdot(\lambda+2) + \lambda + \lceil\log_2|\bbG|\rceil$ bits. The number of PRG invocations in $\Gen$ is at most $2(n+\lceil \frac{\log |\bbG|}{\lambda+2} \rceil)$ and the number of PRG invocations in $\Eval$ is at most $n+\lceil \frac{\log |\bbG|}{\lambda+2} \rceil$.
\end{theorem}

\paragraph{Intuition.}
We now give a high-level description of the tree-based construction, which provides intuition for the more formal description that will follow.  For simplicity, consider first the case of a DPF with a single-bit output $\beta=1$. 

At a high level, each of the two keys defines a GGM-style binary tree~\cite{goldreich1986construct} with $2^n$ leaves, where the leaves are labeled by inputs $x\in\{0,1\}^n$. We will refer to a path from the root to a leaf labeled by $x$ as the {\em evaluation path} of $x$, and to the evaluation path of the special input $\alpha$ as the {\em special evaluation path}. Each node $v$ in a tree will be labeled by a string of length $\lambda+1$, consisting of a {\em control bit} $t$ and a $\lambda$-bit {\em seed} $s$, where the label of each node is fully determined by the label of its parent. The function $\Eval^\pt$ will compute the labels of all nodes on the evaluation path to the input $x$, using the root label as the key, and output the control bit of the leaf.

We would like to maintain the invariant that for each node outside the special path, the two labels (on the two trees) are identical, and for each  node on the special path the two control bits are different and the two seeds are indistinguishable from being random and independent. Note that since the label of a node is determined by that of its parent, if this invariant is met for a node outside the special path then it is automatically maintained by its children. Also, we can easily meet the invariant for the root (which is always on the special path) by just explicitly including the labels in the keys. The challenge is to ensure that the invariant is maintained also when leaving the special path. 

Towards describing the construction, it is convenient to view the two labels of a node as a mod-2 additive secret sharing of its label, consisting of shares $[t]=(t_0,t_1)$ of the control bit $t$ and shares $[s]=(s_0,s_1)$ of the $\lambda$-bit seed $s$.  That is, $t=t_0\oplus t_1$ and $s=s_0\oplus s_1$. The construction employs the same two types of homomorphism of additive secret sharing used in the square-root construction: namely, linear homomorphism that allows for conditional correction and PRG-based homomorphism $[s']=G([s])$ mapping a shared $s=0^\lambda$ to a longer shared $s'=0^m$ and random $s\in\{0,1\}^\lambda$ to a longer pseudorandom $s'$.
To maintain the above invariant along the evaluation path, we use the two types of homomorphism as follows. Suppose that the labels of the $i$-th node $v_i$ on the evaluation path are $[s],[t]$. To compute the labels of the $(i+1)$-th node, the parties start by locally computing $[S]=G([s])$ for a PRG $G:\{0,1\}^\lambda\to\{0,1\}^{2\lambda+2}$, parsing $[S]$ as $[s^L, t^L, s^R, t^R]$. The first two values correspond to labels of the left child and the last two values correspond to labels of the right child. 

To maintain the invariant, the keys will include a correction word $CW$ for each level $i$. As discussed above, we only need to consider the case where $v_i$ is on the special path. By the invariant we have $t=1$, in which case the correction will be applied. Suppose without loss of generality that $\alpha_i=1$. This means that the left child of $v_i$ is off the special path whereas the right child is on the special path. To ensure that the invariant is maintained, we can include in both keys the correction $CW^{(i)}=(s^L, t^L, s^R\oplus s', t^R\oplus 1)$ for a random seed $s'$. Indeed, this ensures that after the correction is applied, the labels of the left and right child are $[0],[0]$ and $[s'],[1]$ as required. But since we do not need to control the value of $s'$, except for making it pseudo-random, we can instead use the correction  $CW^{(i)}=(s^L, t^L, s^L, t^R\oplus 1)$ that can be described using $\lambda+2$ bits.  This corresponds to $s'=s^L\oplus s^R$. The $n$ correction values $CW^{(i)}$ are computed by $\Gen^\pt$ from the root labels by applying the above iterative computation along the special path, and are included in both keys. 

Finally, assuming that $\beta=1$, the output of  $\Eval^\pt$ is just the shares $[t]$ of the leaf corresponding to $x$.  A different value of $\beta$ (from an arbitrary Abelian group) can be handled via an additional correction $CW^{(n+1)}$.

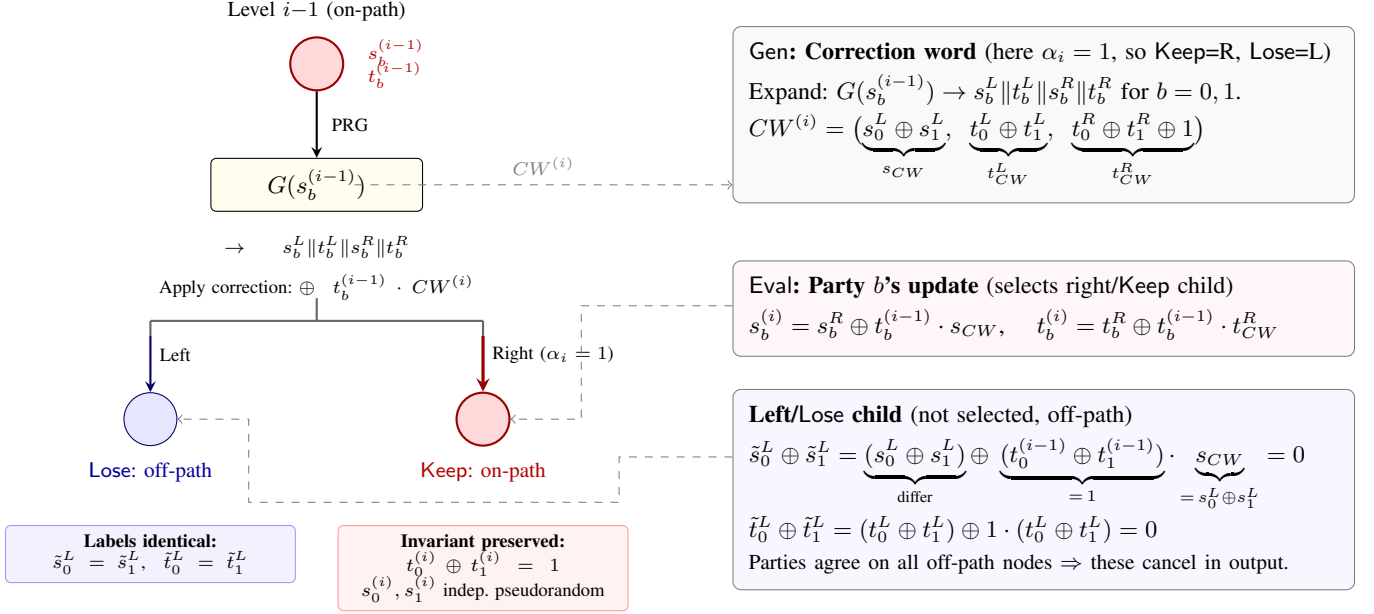
\begin{figure*}[t]
\centering
\begin{tikzpicture}[
  node_style/.style={circle, draw, minimum size=0.7cm, font=\small},
  on_path/.style={node_style, fill=red!15, draw=red!60!black, thick},
  off_path/.style={node_style, fill=blue!10, draw=blue!40!black},
  eqbox/.style={draw=black!50, rounded corners=3pt, fill=gray!5, inner sep=6pt, font=\small},
  arr/.style={->, >=stealth, thick},
  every node/.style={font=\small}
]


\node[on_path] (parent) at (0, 0) {};
\node[font=\footnotesize, anchor=south] at (0, 0.45) {Level $i{-}1$ (on-path)};
\node[font=\scriptsize, red!70!black, anchor=west] at (0.55, 0.15) {$s^{(i-1)}_b$};
\node[font=\scriptsize, red!70!black, anchor=west] at (0.55, -0.15) {$t^{(i-1)}_b$};

\node[draw, rounded corners=2pt, fill=yellow!8, minimum width=2.8cm, minimum height=0.7cm, font=\small] (prg) at (0, -1.6) {$G(s^{(i-1)}_b)$};
\draw[arr] (parent) -- (prg) node[midway, right, font=\scriptsize, xshift=2pt] {PRG};

\node[font=\scriptsize, anchor=north, text width=4.5cm, align=center] (parse) at (0, -2.15) {
$\to\; s^L_b \| t^L_b \| s^R_b \| t^R_b$
};

\node[font=\scriptsize, anchor=north, text width=6cm, align=center] (correct) at (0, -2.65) {
Apply correction: $\oplus\; t^{(i-1)}_b \cdot CW^{(i)}$
};

\draw[thick, black!60] (-2.2, -3.4) -- (2.2, -3.4);
\draw[thick, black!60] (0, -3.1) -- (0, -3.4);
\draw[thick, black!60] (-2.2, -3.4) -- (-2.2, -3.6);
\draw[thick, black!60] (2.2, -3.4) -- (2.2, -3.6);

\node[font=\scriptsize, anchor=west] at (-2.2, -3.85) {Left};
\node[font=\scriptsize, anchor=west] at (2.2, -3.85) {Right ($\alpha_i = 1$)};

\node[off_path] (lose) at (-2.2, -4.7) {};
\node[font=\footnotesize, anchor=north, blue!60!black] at (-2.2, -5.15) {$\Lose$: off-path};

\node[on_path] (keep) at (2.2, -4.7) {};
\node[font=\footnotesize, anchor=north, red!70!black] at (2.2, -5.15) {$\Keep$: on-path};

\draw[arr, blue!40!black] (-2.2, -3.6) -- (lose);
\draw[arr, red!60!black, very thick] (2.2, -3.6) -- (keep);


\node[eqbox, text width=7.8cm, anchor=north west] (genbox) at (5.5, 0.5) {
\textbf{$\Gen$: Correction word} (here $\alpha_i=1$, so $\Keep$=R, $\Lose$=L)\\[4pt]
Expand: $G(s^{(i-1)}_b) \to s^L_b \| t^L_b \| s^R_b \| t^R_b$ for $b=0,1$.\\[3pt]
$CW^{(i)} = \bigl(\underbrace{s^L_0 \oplus s^L_1}_{s_{CW}},\;\; \underbrace{t^L_0 \oplus t^L_1}_{t^L_{CW}},\;\; \underbrace{t^R_0 \oplus t^R_1 \oplus 1}_{t^R_{CW}}\bigr)$
};

\node[eqbox, text width=7.8cm, anchor=north west, fill=red!3] (evalbox) at (5.5, -2.6) {
\textbf{$\Eval$: Party $b$'s update} (selects right/$\Keep$ child)\\[4pt]
$s^{(i)}_b = s^R_b \oplus t^{(i-1)}_b \cdot s_{CW}$, \quad
$t^{(i)}_b = t^R_b \oplus t^{(i-1)}_b \cdot t^R_{CW}$
};

\node[eqbox, text width=7.8cm, anchor=north west, fill=blue!3] (losebox) at (5.5, -4.3) {
\textbf{Left/$\Lose$ child} (not selected, off-path)\\[4pt]
$\tilde{s}^L_0 \oplus \tilde{s}^L_1 = \underbrace{(s^L_0 \oplus s^L_1)}_{\text{differ}} \oplus\; \underbrace{(t^{(i-1)}_0 \oplus t^{(i-1)}_1)}_{=\,1} \cdot \underbrace{s_{CW}}_{=\,s^L_0 \oplus s^L_1} = 0$\\[2pt]
$\tilde{t}^L_0 \oplus \tilde{t}^L_1 = (t^L_0 \oplus t^L_1) \oplus 1 \cdot (t^L_0 \oplus t^L_1) = 0$\\[2pt]
{\footnotesize Parties agree on all off-path nodes $\Rightarrow$ these cancel in output.}
};

\draw[dashed, gray, ->] (0.5, -1.6) -- (genbox.west |- 0,-1.6) node[midway, above, font=\scriptsize, text=gray] {$CW^{(i)}$};
\draw[dashed, gray, ->] (evalbox.west |- 0,-3.2) -- (3.5, -3.2) -- (3.5, -4.7) -- (keep.east);
\draw[dashed, gray, ->] (losebox.west |- 0,-5.2) -- (4.0, -5.2) -- (4.0, -5.8) -- (-0.9, -5.8) -- (-0.9, -4.7) -- (lose.east);


\node[draw=blue!40, rounded corners=2pt, fill=blue!5, font=\scriptsize, text width=3.6cm, align=center, anchor=north] at (-2.2, -6.1) {
\textbf{Labels identical:}\\
$\tilde{s}^L_0 = \tilde{s}^L_1$, \; $\tilde{t}^L_0 = \tilde{t}^L_1$
};

\node[draw=red!40, rounded corners=2pt, fill=red!5, font=\scriptsize, text width=3.6cm, align=center, anchor=north] at (2.2, -6.1) {
\textbf{Invariant preserved:}\\
$t^{(i)}_0 \oplus t^{(i)}_1 = 1$\\
$s^{(i)}_0, s^{(i)}_1$ indep. pseudorandom
};

\end{tikzpicture}

\caption{Illustration of a single level in the tree-based DPF, shown for $\alpha_i = 1$ (right $=$ $\Keep$, left $=$ $\Lose$). The parent on-path node holds seeds $s^{(i-1)}_b$ with $t^{(i-1)}_0 \oplus t^{(i-1)}_1 = 1$. After PRG expansion and conditional correction, the right/$\Keep$ child preserves the invariant (differing control bits, independent seeds), while the left/$\Lose$ child's labels become identical across both parties. The case $\alpha_i = 0$ is symmetric (left $=$ $\Keep$, right $=$ $\Lose$).}
\label{fig:tree-dpf-detail}
\end{figure*}

\paragraph{The formal construction.} A full description of the construction that follows the above blueprint and is used for proving Theorem~\ref{thm:optimized-DPF} is given in Figure~\ref{fig:DPFx4}.
The core idea for how the tree is defined at each level during key generation and evaluation is depicted in Fig.~\ref{fig:tree-dpf-detail}.

\begin{figure}
{\bf Optimized Distributed Point Function $(\Gen^\pt,\Eval^\pt)$}\\
Let $G : \{0,1\}^\lambda \to \{0,1\}^{2(\lambda+1)}$ be a pseudorandom generator.\\
Let $\Convert_\bbG : \{0,1\}^\lambda \to \bbG$ be a map converting a random $\lambda$-bit string to a pseudorandom group element of $\bbG$.
  \vspace{.1in}

$\Gen^\pt(1^\lambda, \alpha,\beta,\bbG)$:
  \begin{algorithmic}[1]
  \State Let $\alpha = \alpha_1,\ldots,\alpha_n \in \{0,1\}^n$ be the bit decomposition of $\alpha$
  \State Sample random $s^{(0)}_0 \gets \{0,1\}^\lambda$ and $s^{(0)}_1 \gets \{0,1\}^\lambda$
  \State Let $t^{(0)}_0 = 0$ and $t^{(0)}_1 = 1$  \label{step:t}
  \For {$i = 1$ to $n$}		
    \State $s^L_0||t^L_0 ~\big|\big|~s^R_0 || t^R_0 \gets G(s^{(i-1)}_0)$ and
    	$s^L_1||t^L_1 ~\big|\big|~s^R_1 || t^R_1 \gets G(s^{(i-1)}_1)$.
    \If {$\alpha_i=0$} $\Keep \gets L$, $\Lose \gets R$
    \Else ~$\Keep \gets R$, $\Lose \gets L$
    \EndIf 
    \State $s_{CW} \gets s^\Lose_0 \oplus s^\Lose_1$
    \State $t^L_{CW} \gets t^L_0 \oplus t^L_1 \oplus \alpha_i \oplus 1$ and
		$t^R_{CW} \gets t^R_0 \oplus t^R_1 \oplus \alpha_i$
    \State $CW^{(i)} \gets s_{CW} || t^L_{CW} || t^R_{CW}$
    \State $s^{(i)}_b \gets s^\Keep_b \oplus t^{(i-1)}_b \cdot s_{CW}$ for $b = 0,1$
    \State $t^{(i)}_b \gets t^\Keep_b \oplus t^{(i-1)}_b \cdot t^\Keep_{CW}$ for $b=0,1$
  \EndFor
  \State $CW^{(n+1)} \gets (-1)^{t^{n}_1}\cdot\big[\beta - \Convert(s^{(n)}_0) + \Convert(s^{(n)}_1)\big]\in\bbG$
  \State Let $k_b = s^{(0)}_b || CW^{(1)} || \cdots || CW^{(n+1)}$ \\
  \Return $(k_0,k_1)$
  \end{algorithmic}
  
  \vspace{.1in}

$\Eval^\pt(b, k_b, x)$:
  \begin{algorithmic}[1]
  \State Parse $k_b = s^{(0)} || CW^{(1)} || \cdots || CW^{(n+1)}$, and let $t^{(0)} = b$.
  \For {$i=1$ to $n$}
    \State Parse $CW^{(i)} = s_{CW} || t^L_{CW} || t^R_{CW}$
    \State $\tau^{(i)} \gets G(s^{(i-1)}) \oplus (t^{(i-1)} \cdot \big[s_{CW} || t^L_{CW} || s_{CW} || t^R_{CW} \big])$
    \State Parse $\tau^{(i)} = s^L || t^L ~\big|\big|~ s^R || t^R  \in \{0,1\}^{2(\lambda+1)}$
    \If {$x_i = 0$}  $s^{(i)} \gets s^L, t^{(i)} \gets t^L$ 
    \Else ~$s^{(i)} \gets s^R$, $t^{(i)} \gets t^R$
    \EndIf
  \EndFor\\
  \Return $(-1)^b\cdot\big[\Convert(s^{(n)}) + t^{(n)} \cdot CW^{(n+1)}\big] \in \bbG$ 
  \end{algorithmic}

\caption{Pseudocode for optimized DPF construction for the class $f_{\alpha,\beta} : \{0,1\}^n \to \bbG$. The symbol $||$ denotes string concatenation. Subscripts 0 and 1 refer to party id. All $s$ values are $\lambda$-bit strings and $t$ values are a single bit.}
\label{fig:DPFx4}
\end{figure}

\subsection{Multi-Party DPF}
\label{sec:multi-partyDPF}


A natural generalization of two-party DPF is multi-party DPF, in which a dealer distributes the keys to $m$ parties with $1 \le t<m$ corruptions. We refer to schemes with these parameters as $(m,t)$-DPF. Multi-party DPF schemes were constructed in three regimes for the number of corrupt parties.

The first regime is $t=1$, and the performance measures of known schemes \cite{bunn2022cnf,C:ABG+24} in this regime roughly match the performance of the $2$-party scheme in Figure \ref{fig:DPFx4}. These multi-party schemes have the additional benefit that the output is shared in a {\em multiplicative} secret sharing, allowing the computation of a circuit with low multiplicative depth on the output without interaction. In the second regime, $t=m-1$, i.e., the adversary controls all but one of the parties. In this regime, the best known schemes with security from one-way functions \cite{EC:BoyGilIsh15,goel2025multiparty} are similar to the square root scheme in Section \ref{sec:construct} with key size and evaluation time that are proportional to a square root of the domain size. In the final regime, $1 < t <m/2$, or more precisely $m=td+1$ for $d>1$. In this regime, the most efficient scheme \cite{bunn2022cnf} composes a DPF for full threshold with an information-theoretically secure DPF to achieve key size and evaluation time that are proportional to $2^{n/(2d)}$.

\paragraph{Multiplicative DPF.}
The following theorem summarizes the current state of the art for $(m,1)$-DPF, in which the output is shared in a {\em multiplicative sharing}, e.g.\ CNF secret sharing \cite{ito1989secret} or Shamir secret sharing \cite{shamir1979share}. This type of secret sharing enables computing a polynomial of degree $m-1$ on the secret by manipulating the secret shares locally without interaction.

\begin{theorem}
    \label{th:mulDPF}
Suppose 
$G$ is a pseudorandom generator with seed length $\lambda$, let $\bbG$ be a group and let the number of parties be $m>2$.
Then, there exists an  $(m,1)$ DPF scheme for the family of point functions $f_{\alpha,\beta} : \{0,1\}^n \to \bbG$ with output that is a CNF secret shared over $\bbG$ with threshold $1$ and key size per party that is $O(mn(\lambda+\log m)+m^2 \log |\bbG|)$.
\end{theorem}

When only additive reconstruction is needed, it suffices for $\bbG$ to be an
Abelian group.  When the CNF-shared output is used for multiplication, e.g., via conversion into Shamir shares~\cite{cramer2005share}, the group $\bbG$ should
be viewed as the additive group of a commutative ring, and the local products are taken in that ring.

A scheme that satisfies Theorem \ref{th:mulDPF} is constructed in \cite{C:ABG+24}. This scheme follows the general tree structure of the two-party scheme from Section \ref{sec:tree} with two main differences. Each level of the tree is associated with roughly $m$ different correction words and a node in the tree holds both a seed for the PRG and an ID of one of the correction words. The evaluation procedure ensures that in any node of the tree, which is {\em not} on the path to $\alpha$, the seeds and IDs of all $m$ parties are identical. In any node of the tree, which is on the path to $\alpha$, the seeds are drawn independently from a pseudorandom distribution and all the IDs of correction words are distinct. Maintaining this invariant in the leaves ensures that by stretching the final seeds to a length of $m-1$ group elements in $\bbG$, and with correction words of appropriate size, the parties can output shares in a replicated, or CNF, secret sharing scheme of $\beta$ in the point $\alpha$ and of $0$ in every other point $\alpha \ne x \in \{0,1\}^n$. If $\bbG$ is a ring then CNF secret sharing enables non-interactive computation of shares of polynomials of degree at most $m-1$ over the ring. If $\bbG$ is a field $\F, |\F|>m$, then there is a procedure to first locally transform the CNF shares into Shamir shares~\cite{cramer2005share} and then compute polynomials of degree at most $m-1$ over $\F$.

\paragraph{Multi-party DPF with many corruptions.} In the regime of $t>1$, the schemes are significantly less efficient. The reason for this, which was already mentioned in Section \ref{sec:warmup}, is that two-party constructions encode a zero bit by parties holding identical seeds. Obtaining multiple encoded zero bits from a singled encoded zero bit is possible non-interactively by locally stretching the identical seeds to longer identical strings. However, in the multi-party setting, one cannot take such a direct approach with identical seeds, as an adversary that controls two or more parties can determine if they share a zero bit. 

In the full threshold regime, i.e. $t=m-1$, the two state of the art constructions \cite{EC:BoyGilIsh15,goel2025multiparty} achieve a square-root scheme, improving over the trivial information-theoretic approach, but leaving a large gap in performance between $2$-party and multi-party schemes. Both constructions also place restrictions on the output group. Theorem \ref{th:multBGI} summarizes the results of the multi-party scheme from \cite{EC:BoyGilIsh15}.

\begin{theorem}
    \label{th:multBGI}
Suppose $G: \{0,1\}^\lambda \to \{0,1\}^{2\lambda}$ is a pseudorandom generator, let $\bbG=\F_2$ and let the number of parties be $m>2$.
Then, there exists an  $(m,m-1)$ DPF scheme for the family of point functions $f_{\alpha,\beta} : \{0,1\}^n \to \bbG$ with output that is a additively shared over $\bbG$ and key size per party that is $2^{\left \lceil n/2 \right\rceil} \cdot 2^{\left \lceil (m-2)/2 \right\rceil} \lambda  (1+o(1))$. 
\end{theorem}

The general structure of the construction is similar to the square-root DPF in Section \ref{sec:warmup}, where the domain $\zo^n$ is viewed as a matrix of roughly equal dimensions. There exist a unique row $\alpha_1$ and  a unique column $\alpha_2$ in this matrix that correspond to the special input $\alpha \in \zo^n$. The main difference between the two schemes is in the distribution pattern of the PRG seeds. The $m$-party scheme of \cite{EC:BoyGilIsh15} samples $2^{m-1}$ independent seeds for each row of the matrix that represents the domain $\{0,1\}^n$. In every row that {\em does not} contain $\alpha$, the $\Gen$ procedure selects one seed $s_S$ for every subset $S \subseteq [m]$ of {\em even} size and provides it to all parties $i \in S$. In $\alpha_1$, the {\em critical} row that contains $\alpha$, $\Gen$ provides one fresh seed $s_{S'}$ for every subset $S' \subseteq [m]$ of parties of {\em odd} size. $\Eval$ expands every seed to the length of the row. In all non-critical rows the sum of expanded strings over $\bbG=\F_2$ is $0$, while in the critical row the sum of strings is some pseudo random $v$. $\Gen$ also chooses a correction word $cw$ such that $v \oplus cw=e_{\alpha_2}$, i.e. a unit vector with $1$ at $\alpha_2$ and zero anywhere else. $\Gen$ also provides an additive secret sharing over $\F_2$ of a control bit per row, which is $1$ in $\alpha_1$ and $0$ anywhere else. $\Eval$ completes its operation after expanding the seeds and computing the sum of all the expanded strings by adding the product of the public correction word and the share of the control bit associated with the evaluated row. 

Security is ensured due to two properties. First, the view of the distribution of the seeds, given $m-1$ of the parties, is identical for the row that contains $\alpha$ and for every other row. Second, the party that is not corrupted by the adversary receives one seed that no other party has. Therefore, since the value of the correction word is $cw=v \oplus  e_{\alpha_2}$, and $v$ is the sum of all the expanded seeds, the correction word is  computationally indistinguishable from random for the adversary.

The above description slightly differs from the scheme in \cite{EC:BoyGilIsh15} since it uses a single correction word instead of $2^{m-1}$ correction words. This allows, through a careful balancing of the number of rows and columns of the domain matrix, to reduce the dependence of the key size on the number of parties from about $2^m$ to roughly $2^{m/2}$.

\begin{remark}
    The output group can be generalized to $\mathbb{Z}_p$ for any small prime $p$ at the cost of increasing the key size by a multiplicative factor of $O \left((p/2)^{m/2} \right)$. Generalization of the output group to $\mathbb{Z}_k$ for a smooth integer $k$ with distinct prime factors is straightforward via the Chinese Remainder Theorem. 
\end{remark}

The authors of \cite{goel2025multiparty} set out to improve the exponential dependence in key size (and evaluation time) on the number of parties in the above construction. The following theorem presents the main result.

\begin{theorem}
    \label{th:multGWW}
Suppose $G: \{0,1\}^\lambda \to \{0,1\}^{2\lambda}$ is a pseudorandom generator, let $\bbG=\mathbb{Z}_{p^q}$ for a prime $p$ and integer $q$ and let the number of parties be $m>2$.
Then, there exists an  $(m,m-1)$ DPF scheme for the family of point functions $f_{\alpha,\beta} : \{0,1\}^n \to \bbG$ with output that is additively shared over $\bbG$ and key size per party that is $\mbox{poly}(m,\lambda) \cdot 2^{n/2}$.
\end{theorem}

This work first shows that any seed distribution pattern which is {\em deterministic} and has the following three properties must have exponential size: (1) Any seed in a non-critical row must be used an even number of times by all the parties together, (2) Each party receives a unique seed for the critical row, and (3) The distribution of seeds is identical between the critical row and other rows. As a corollary, the seed distribution scheme in \cite{EC:BoyGilIsh15} is optimal if these properties are met. On the other hand, \cite{goel2025multiparty} shows that a {\em randomized} pattern for seed distribution can have a polynomial number of seeds per party. In more detail, $\Gen$ distributes the same seed to $a$ random pairs of parties (including a pair of seeds to the same party) for every row. $\Gen$ distributes $b$ additional seeds, one seed per party, in the critical row. This type of approach satisfies the first two desired properties of the seed distribution. By a careful choice of $a,b$, the statistical distance between the distribution of the critical row and a non-critical row is inverse-polynomial (although non-negligible). The final step is a local privacy amplification step proposed in \cite{boyle2022programmable} via Locally Decodable Codes, achieving negligible distinguishing advantage. Other components of the construction, such as the correction word and the control bits, are copied from \cite{EC:BoyGilIsh15} without impacting the asymptotic key size.

\paragraph{Leveraging multiplicativity for better DPF key size.}
In the final regime, $m=td+1$ for some $d>1$. \cite{bunn2022cnf} proposes a construction in this regime with properties that are summarized in Theorem \ref{th:multBunn}. 
\begin{theorem}   \label{th:multBunn}
Suppose $G: \{0,1\}^\lambda \to \{0,1\}^{2\lambda}$ is a pseudorandom generator, let $\bbG=\F_2$, let $d,t>1$ and let the number of parties be $m=dt+1$.
Then, there exists an  $(m,t)$ DPF scheme for the family of point functions $f_{\alpha,\beta} : \{0,1\}^n \to \bbG$ with output that is additively shared over $\bbG$ and key size per party that is $O \left(2^{n/2d} \cdot 2^{m^t/2}\cdot m^t d\lambda \right)$.
\end{theorem}

The scheme uses two ideas: the DPF scheme with security against all but one corrupt parties from \cite{EC:BoyGilIsh15} and the multiplicative property of CNF secret sharing. In more detail, the scheme views the domain $\{0,1\}^n$ as a $d$-dimensional cube with each side of length $2^{n/d}$. If $\alpha=(\alpha_1,\ldots,\alpha_d)$ in this cube, then the goal is to compute $f_{\alpha_1,\beta} \cdot \prod_{i=2}^d f_{\alpha_i,1}$. Let $(\Gen',\Eval')$ be the DPF scheme for $m'=\binom{m}{t}$ parties with threshold $t'=m'-1$. For each point function $f_{\alpha_i,\cdot}$, $\Gen$ runs $\Gen'(\alpha_i,\cdot)$. Each of the $m'$  ``virtual'' parties is associated with a subset of size $t$ out of the $m$ real parties. A real party receives all the keys of subsets to which it {\em does not} belong. Therefore, a subset of $t$ real parties misses one key, ensuring security. $\Eval(x)$ works in two steps. If the coordinates of $x$ are $(x_1,\ldots,x_d)$, then $\Eval$ first runs $\Eval'$ on every $x_i$ that the party holds. Now, the parties hold secret shares of $f_{\alpha_i,\cdot}(x_i)$ for $i=1,\ldots,d$. More precisely, the parties hold $(m',d)$ CNF shares of each $f_{\alpha_i,\cdot}(x_i)$. In the second step of $\Eval$, the parties exploit the multiplicative property of CNF sharing to non-interactively compute additive shares of $f_{\alpha_1,\beta} \cdot \prod_{i=2}^d f_{\alpha_i,1}$, which is $\beta$ if $x=\alpha$ and is $0$ if $x \ne \alpha$.  


\subsection{Extending to Distributed Comparison Functions (DCF)}
\label{sec:DCF}

Recall that a DCF is an FSS for the class of {\em comparison} functions $f_{\alpha, \beta}^{<}: \zo^n \to \G$, where $\alpha \in \zo^n$, $\beta \in \G$, and $f_{\alpha, \beta}^{<}(x)= \beta$ for $x < \alpha$ and $0 \in \G$ for $x\ge \alpha$. Here we view $x=x_1\ldots x_n$ and $\alpha=\alpha_1\ldots\alpha_n$ as representing $n$-bit integers, where $x_1$ and $\alpha_1$ are the most significant bits. 

A natural question is whether we can use {\em any} DPF construction to obtain a DCF. This turns out to be indeed possible, though with a multiplicative overhead of $n$. Indeed, $x<\alpha$ if and only if (1) $x_1<\alpha_1$ or (2) $x_1=\alpha_1$ and $x_2<\alpha_2$ or (3) $x_1x_2=\alpha_1\alpha_2$  and $x_3<\alpha_3$, ..., or $(n)$:  $x_1x_2\ldots x_{n-1}=\alpha_1\alpha_2\ldots\alpha_{n-1}$  and $x_n<\alpha_n$. Rewriting condition $(i)$ as $x_1x_2\ldots x_{i}1=
\alpha_1\alpha_2\ldots\alpha_{i-1}0\alpha_{i}$, and noticing that the $n$ conditions are mutually exclusive, we can write 
\begin{equation}
\label{eq-dcf}
 f_{\alpha, \beta}^{<}(x)=\sum_{i=1}^{n}f_{\alpha^i, \beta}(x^i)  
\end{equation}
 for $\alpha^i=\alpha_1\alpha_2\ldots\alpha_{i-1}0\alpha_{i}$ and $x^i=x_1x_2\ldots x_{i}1$. This generically reduces DCF to $n$ instances of DPF.

Can we do better? It turns out that the tree-based construction of DPF from Section~\ref{sec:tree} can be converted into a DCF with a very small overhead. In fact, the DPF key for $f_{\alpha,\beta}$ can also serve as a DCF key for $f_{\alpha, \beta}^{<}$ in the useful case where $\G=\Z_2$ and $\beta=1$. This follows from the fact that the tree-based DPF construction effectively generates not only additive shares of $f_{\alpha,\beta}$, but also additive shares of $f_{\alpha',\beta}$ for every prefix $\alpha'$ of $\alpha$. This implies that each term $f_{\alpha^i, \beta}(x^i)$ from Eq.~(\ref{eq-dcf}) can be evaluated locally from the DPF keys: If $x_i=1$, the term is 0 can be omitted from the summation. If $x_i=0$, then we can write this term as $f_{\alpha',\beta}(x_1\ldots x_{i-1}1)$ where $\alpha'$ is the length-$i$ prefix of $\alpha$. 

To handle the case of a general output group $\G$ and payload $\beta$, it suffices to augment the tree-based DPF so that every intermediate level is also treated as the final output level, yielding the prefix-DPF instances $f_{\alpha',\beta}$ required by the above construction. This involves roughly $n\cdot\log|\G|$ extra correction bits in each key.

Finally, we note that two instances of DCF can be used for a {\em distributed interval function}, which evaluates to $\beta$ on a secret interval $[a,b]$ and to 0 elsewhere. 

\subsection{Extending to Distributed Multi-Point Functions (DMPF)}
\label{sec:DMPF}

Many applications of DPF, such as their use within constructions of efficient pseudorandom correlation generators discussed in Section~\ref{sec:PCG}, actually require a compressed secret sharing of a (sparse) weight-$t$ vector.  This corresponds to a distributed {\em multi}-point function (DMPF), where the secret function $f_{A,B}: \zo^n \to \bbG$ has up to $t$ nonzero outputs for some public parameter $t$ 
(see notation in Section~\ref{sec:extensions}).

In what follows, we provide a brief overview of existing DMPF constructions. 


\begin{itemize}

  \item {\em Na\"{i}ve DMPF.} In its simplest form, an additively shared vector of weight $\le t$ can be obtained by adding together $t$ additively shared vectors of weight $\le 1$. In turn, one can build $t$-DMPF via $t$ independent copies of standard DPF, resulting in $\times t$ cost in all metrics.

   \item {\em ``Big-state'' DMPF}~\cite{boyle2025improved}. One can generalize the indicator bit in the standard tree-based DPF construction from~\cite{CCS:BoyGilIsh16} (Section~\ref{sec:tree}) to an indicator $t$-bit {\em string}, which identifies multiple nonzero entries of the shared vector. The construction maintains the invariant that for a node in the tree lying on the $k$th accepting path in the depth-$i$ level, its indicator string
   is the one-hot encoding of $k$, namely
   $e_k = 0^{k-1}||1||0^{t-k} \in \zo^t$.  
   This enables the parties to (unknowingly) perform different conditional corrections for $t$ positions in each layer.
   
   To achieve this, each level of the tree will contain $t$ sets of correction words, each containing $\lambda + 2t$ bits. Each set will be used in an analogous manner to the standard tree-based construction. Namely, at each node of evaluation in the tree, and for every $\ell \in [t]$, each party will conditionally XOR in the $\ell$th correction word information, based on the value of the $\ell$th bit of its local indicator string.
   
   The costs of the big-state DMPF scheme grow quickly (quadratically) with the vector weight $t$; however, this construction typically outperforms the full-domain evaluation costs of competing approaches for small weights in the range $3\le t \le70$.

  \item {\em DMPF from batch codes.} DMPF construction with better asymptotic efficiency can be obtained from a {\em probabilistic batch code} (PBC)~\cite{AngelCLS18,CCS:BoyleCGI18,CCS:SchoppmannGR019,EC:CastroP22,bombar2024foleage,ARR25}.
A batch code~\cite{IshaiKOS04, PatersonSW09} is a distributed encoding scheme that allows a large database to be distributed across $\ell$ servers such that any ``batch'' of $t$ database symbols can be retrieved by reading one symbol from each server. 
The goal is to minimize the total storage requirements, using $\ell>t$ servers to beat the simple solution of duplicating the database across all servers. 

Here we consider PBC, a probabilistic variant of batch codes, where the encoding replicates each database symbol among a set of servers using a probabilistic replication pattern. Efficient constructions of PBC can be based on cuckoo hashing~\cite{AngelCLS18,CCS:BoyleCGI18,CCS:SchoppmannGR019,Yeo23}. 

When applied to FSS for multi-point functions, a PBC 
reduces the task of building a weight-$t$ DMPF on domain size $2^n = N$ (the large database size), to $\ell > t$ separate {\em weight-$1$}  DPF instances, each with domain size $<N$.  The batch code provides a mapping between positions of the $\ell$ short DPF vectors to positions of the original length-$N$ vector, so that each of the $N$ positions is covered at least once, and the following combinatorial property is satisfied: For any subset of $t$ positions within the length $N$ vector, there exists a choice of (at most) {\em one} entry from each of the $\ell$ vectors, which covers these $t$ items. The key size of the DMPF scales with the number of PBC servers $\ell$, and the full evaluation time scales with the total storage. The latter can get close to $2N$ in a cuckoo hashing based PBC.




    \item {\em OKVS-based DMPF.}
  The final class of DMPF constructions we discuss rely on {\em oblivious key-value stores} (OKVS), originally proposed as a data structure supporting efficient private set intersection (PSI) protocols \cite{C:GarimellaPRTY21}. Improved OKVS constructions were given in~\cite{CCS:RaghuramanR22,USENIX:BienstockPSY23}. OKVS can be used as a method to encode specific positions in the binary tree (the ``keys'') paired up with specific pieces of information needed for correction (the ``values''), and these encodings can be given in the DMPF keys. An OKVS abstracts and generalizes the concept of polynomial interpolation, enabling higher concrete efficiency.
  
  The key size, key generation time and evaluation time of the corresponding OKVS-based DMPF are closely related to the OKVS instantiation. A simple polynomial-based OKVS can be used to provide improvements over na\"{i}ve DMPF~\cite{boyle2025improved,SLAMPFSS}.
  By applying the ``random bands'' RB-OKVS construction of~\cite{USENIX:BienstockPSY23}, one obtains a DMPF scheme with fastest evaluation time for a wide range of practically useful parameters~\cite{boyle2025improved}.   

  
\end{itemize}

\paragraph{Comparison of approaches.}
The best choice of DMPF varies strongly based on parameter regimes, with each of the four constructions outperforming others within some region. 
For example, based on experimental results from~\cite{boyle2025improved}, the optimal DMPF construction in terms of minimal $\fulleval$ time is largely independent of the choice of the domain size $2^n = N$, and varies for different ranges of $t$ as follows: 

\begin{itemize}
\item Na\"{i}ve DMPF: For $t\leq 2$
\item Big-state DMPF: For $3\leq t \leq 70$
\item OKVS-based DMPF: For $70<t < 10^4$
\item PBC-based DMPF: Comparable to OKVS starting at $t\ge 10^4$, and best starting around $t\approx$ 21,000. 
\end{itemize}
For the case of single-evaluation $\Eval$, the picture is similar, but with an earlier switch between the big-state and OKVS (roughly $t=9$ in the place of $t=70$).

We remark that each of the dominating regimes 
constitute an important range for applications.  For example, the big-state DMPF is a good fit for use in state-of-the-art PCG implementations \cite{bombar2024foleage} and other applications of DMPF that require small values of $t$.  
OKVS-based and PBC-based DMPF constructions provide improved running times for applications such as unbalanced PSI in the 2-server model, considered in~\cite{DittmerILOEKSS22}. Here a client wants to find the intersection of its set of $t$ keywords (each of length $n$) with a potentially big (or even ``streaming'') set of keywords held by two remote servers.

\subsection{Information-Theoretic DPF}
\label{sec:itdpf}

Nontrivial $m$-party DPF constructions are possible even in the {\em information-theoretic} setting, where each $t$ keys perfectly hide the point function, provided that security threshold $t$ satisfies $t<m/2$.

Such DPF constructions are closely related to protocols for information-theoretic private information retrieval (PIR)~\cite{CGKS95}. The first generation of such schemes were based on Reed-Muller codes, and could achieve 
communication complexity $N^{1/\Theta(m)}$. As it turns out, for $m \ge 3$ these schemes 
imply information-theoretic DPF schemes with similar key size. 

\begin{theorem}[Reed-Muller DPF, implicit in~\cite{CGKS95}]
Let $p \ge 2$ be a prime and $m \ge 2$ an integer. There exists a perfectly secure 
$m$-party DPF, for point functions with input domain $[N]$ and output group $\mathbb{Z}_p$, where each key is of size 
$O(m\log m\log(p) \cdot N^{1/(m-1)})$.
\end{theorem}

The above can be extended to larger security thresholds $t\ge 2$, scaling $m$ by roughly a factor of $t$.

Subsequent information-theoretic PIR schemes~\cite{Yek07,Efr09,DvirGopi15,AlonBeimelLasri25} improved the communication complexity to $N^{o(1)}$ with as few as 2 servers. These 2-server PIR schemes have the two servers output additive shares of a vector of length $N^{o(1)}$, whose inner product with a secret linear combination vector known to the PIR client yields the output. By doubling the number of servers, we can ensure that each answer is known to two servers, and have the client share its linear combination vector between each pair of servers. This results in a 4-server PIR scheme with additive reconstruction, which given known 2-server PIR schemes yields the following.

\begin{theorem}[Perfectly secure 4-DPF~\cite{BGIK22}]
Let $p\ge 3$ be prime and $s\ge 1$. There is a perfectly secure $4$-party DPF ($t=1$) with output group $\Z_{p^{s}}$, domain size $N$, and key size
\[
 O\!\left( s\log(p)\cdot 2^{\,2p\sqrt{\log N\,\log\log N}} \right).
\]
\end{theorem}

Obtaining a 3-party information-theoretic DPF with $N^{o(1)}$ key size is more delicate, and only statistical (rather than perfect) security is known to be achievable. The starting point is a relaxed flavor of 3-party DPF obtained via 3-server PIR constructions from~\cite{Efr09,BIKO12}.
It is relaxed in the sense that there is no way to fully control the nonzero output value $\beta$, and moreover this value inherently depends on $\alpha$. An unfortunate byproduct is that revealing to the parties a ``correction'' that maps that shared output $\hat\beta$ to the target $\beta$ would compromise the secrecy of $\alpha$. 

The high-level idea starts with a slight modification of this relaxed DPF so that the $\beta$ values have at least one bit of entropy even when conditioned on $\alpha$. Then, applying $O(\sigma)$ independent copies, the $\beta$ value has $O(\sigma)$ bits of entropy. Finally, the random $\hat\beta$ is mapped to the target $\beta$ using a compressive linear hash function which is made public, namely part of all 3 keys. The statistical security of this construction follows from the Leftover Hash Lemma~\cite{ILL}.

\begin{theorem}[Statistically secure 3-party DPF]
There is a statistically $2^{-\sigma}$-secure $3$-party DPF for point
functions with output group $\Z_p$ (prime $p$), domain size $N$, and key size
\[
  O\!\left(\sigma\log(p)\cdot 2^{\,k(p)\sqrt{\log N\,\log\log N}}\right),
\]
where $k(2)=6$, $k(3)=10$, and $k(p)=2p$ for $p\ge 5$.
\end{theorem}

Extensions to more general output groups appear in~\cite{li2025efficient}.

\subsection{Overview of State-of-the-Art FSS Constructions}
\label{sec:FSSconstruct}

As of the writing of this article (February 2026), the collection of known FSS constructions are as follows. The given complexity measures are with respect to $n$-bit inputs.
Unless otherwise specified, the results of this section are for $m=2$ parties.

\setcounter{subsubsection}{-1}

\medskip
\subsubsection{``Lowest End'' -- Information-Theoretic FSS}

Perfectly secure $m$-party FSS for linear functions can be implemented by additive secret sharing~\cite{Ben86}.
Another simple and perfectly secure FSS construction that applies to an {\em arbitrary} function class $\{f: \zo^n \to \G\}$ is one that simply secret-shares the truth table of $f$. However, this requires exponential key size $2^n \cdot \log |\G|$.

Yet another simple and perfect FSS includes secret linear combinations of public functions: namely, $f_{\vec \alpha} (x) = \sum_{i \in S} \alpha_i g_i(x)$, where the functions $\{g_i : \zo^n \to \G\}_{i \in S}$ are public for some module $\G$ over ring $R$, and $\vec \alpha = (\alpha_i)_{i \in S}$ are secret $R$-coefficients. Perfectly secure $m$-party FSS for this class can be achieved by simply providing additive secret shares of the linear coefficients $\alpha_i$, with key size $|S| \lceil \log|R| \rceil$~\cite{EC:BoyGilIsh15}. To evaluate the FSS key on input $x$, a party computes each $g_i(x)$ locally in the clear, and outputs the linear combination dictated by its shares of $\alpha_i$. This captures, for example, FSS for low-degree multivariate polynomials, by viewing the $g_i$ as the set of all monomials within the dictated degree.

More sophisticated information theoretic FSS schemes can be obtained for $m>2$ parties, provided that the security threshold $t$ satisfies $t<m/2$. Such constructions will be discussed  in Section~\ref{sec:itdpf}.

\medskip

\subsubsection{``Low End'' -- FSS from One-Way Functions}
For simple but useful function classes (which are the focus of this survey), FSS can be constructed assuming only the existence of one-way functions, or equivalently, of a pseudorandom generator (PRG).  This assumption can be shown to be minimal for $m=2$ parties with nontrivial function classes and FSS key size; see, e.g.,~\cite{EC:GilboaIshai14,EC:BoyGilIsh15}. 

The following constructions all make a black-box use of an arbitrary PRG stretching a $\lambda$-bit random seed to a pseudorandom output of length $\approx 2\lambda$. 
In an AES-based implementation, we typically have $\lambda=128$. 

  \begin{itemize}
  
  \item {\bf Point functions} (DPF).
     The class of point functions consists of functions of the form $f_{\alpha,\beta}$, such that $f_{\alpha,\beta}(x)$ outputs $\beta$ if $x=\alpha$ and 0 otherwise. Here $x,\alpha\in\zo^n$ and $\beta\in\G$ for an Abelian group $\G$. We let $|\beta|$ denote the bit-length of a representation of a group element. When $|\beta|$ is omitted it is understood to be 1. A {\em distributed point function} (DPF) is an FSS scheme for the class of point functions.
    \begin{itemize}
    \item The first nontrivial (2-party) DPF was implicitly constructed in~\cite{ChorG97} in the context of computationally private information retrieval. The key size of this construction is $2^{O(\sqrt{n})}\cdot \lambda$, which is not polynomial in the input length $n$ but still a super-polynomial improvement over the naive solution of additively sharing the size-$2^n$ truth table. 
    \item The notion of DPF was first explicitly introduced in~\cite{EC:GilboaIshai14}, which also gave a recursive construction with key size $O(n^{\log_23}\cdot \lambda)$. This was improved to $O(n \lambda)$ bits in~\cite{EC:BoyGilIsh15} via a tree-based construction.
    \item The best construction to date remains that of Boyle, Gilboa, and Ishai from 2016~\cite{CCS:BoyGilIsh16}, with key size $\approx n\lambda+|\beta|$. More precisely, the key size is $\lambda + n(\lambda+2) - \lfloor \log (\lambda / |\beta|) \rfloor$ bits.\footnote{In particular: $\lambda + n(\lambda + 2)$ for $\lambda$-bit outputs, and $\lambda + n(\lambda+2) - \lfloor\log \lambda \rfloor$ for 1-bit outputs.} Any nontrivial improvement of this construction remains a highly motivated open problem.
    \end{itemize}
    
  \item {\bf Multi-party DPF}.
The complexity of existing $m$-party DPF constructions depends greatly on the desired security threshold $1 \le t \le m-1$.

    \begin{itemize}
    \item For $m>2$ parties, with security against $t=m-1$: The construction presented in~\cite{EC:BoyGilIsh15} has key size $O(2^m 2^{n/2} \cdot\lambda)$ bits. For a small number of parties $m$, this gives a near-quadratic improvement over the naive solution of secret-sharing the truth-table. 
    In \cite{goel2025multiparty}, the dependence on the number of parties was improved from exponential to polynomial via a randomized design, achieving key size 
     $2^{n/2}\cdot{\rm poly}(\lambda,m)$ (for small values of $m$, the construction from~\cite{EC:BoyGilIsh15} has better concrete efficiency).

    The question of improving this square-root bound based on symmetric cryptography alone is one of the central open questions in the area. For example, it is even open a one-way function implies a 3-server DPF secure against 2 corruptions with key size 
    $O(\lambda\cdot 2^{(1/2-\epsilon)n})$, for some constant $\epsilon>0$.
    
    \item For $m>2$ parties with a gap in the security threshold, i.e.\ $1 \le t < m-1$ corruptions: Better DPF constructions can be obtained from information-theoretic private information retrieval schemes~\cite{CGKS95} or from PRG-based DPF~\cite{bunn2022cnf}. 
    
    One can also achieve a stronger form of {\em multiplicative DPF}~\cite{bunn2022cnf,C:ABG+24} in this setting, where $\eval$ outputs multiplicative secret shares (e.g., ``replicated shares'' or ``Shamir shares'') instead of additive shares.  This allows parties to locally convert their $\eval$ outputs to additive secret shares of the {\em product} of up to 
    $\lfloor (m-1)/t\rfloor$ DPF outputs. (For example, for $t=1$, the product of $m-1$ outputs.)

    \end{itemize}

    See Section~\ref{sec:multi-partyDPF} for more detailed discussion of multi-party DPF with $m > 2$ parties.


  \item {\bf Comparison and interval functions.}
    The class of comparison functions consists of functions $f_b$ which output 1 on inputs $x$ with $x < b$, where the input space $\zo^n$ is interpreted as integers from 0 to $2^n-1$. Interval functions $f_{(a,b)}$ output 1 precisely for inputs $x$ that lie within the interval $a < x < b$, and $0$ otherwise. Constructions of FSS for such functions follow a similar structure as DPFs. The best key size for comparison function is comparable to a DPF and for interval functions it is roughly twice the size~\cite{EC:BoyGilIsh15,CCS:BoyGilIsh16,EC:BCG+21}. See Section~\ref{sec:DCF} for the case of {\em distributed comparison functions} (DCF).

  \item {\bf Multi-point functions}.
A distributed {\em multi}-point function (DMPF) is an extension of DPF, where the secret function $f_{A,B}: \zo^n \to \bbG$ has up to $t$ nonzero evaluation points for some publicly known parameter $t$ 
  (see notation in Section~\ref{sec:extensions}).  Many of the applications of DPF can benefit from this extension. For example, the application to PIR by keywords described in Section~\ref{sec:keywords} can be extended using DMPF compute the intersection of a small set of $t$ keywords held by a client with a database $D$ of keywords held by the server, where the communication cost scales linearly with $t$ and logarithmically with $|D|$. 
  
  A simple way of realizing DMPF is by expressing it as the sum of $t$ instances of DPF. The main disadvantage of this approach is that it increases the computational cost of full-domain evaluation by a factor of $t$. In Section~\ref{sec:DMPF} we discuss different approaches for minimizing this computational overhead at the price of a slightly bigger increase to the key size.

  
  \item {\bf Offset function classes.}
The application of FSS to secure computation with preprocessing, proposed in~\cite{boyle2019secure}, provides a low-communication reduction from secure evaluation of a gate $g:\G_{\sf{in}}\to\G_{\sf{out}}$ (mapping additive shares of the input to additive shares of the output), to FSS for a related {\em offset class} which includes, for any $r\in\G_{\sf{in}}$, the function $g_r(x)=g(x-r)$.

It turns out that for many useful nonlinear gate functions $g$, the induced offset class is FSS-friendly in the sense that it admits a concretely efficient PRG-based FSS scheme. As a simple example, if $g$ is a zero-test gate which outputs $1\in\Z_2$ if the input is zero and outputs $0\in\Z_2$ otherwise, then the induced offset class is just the set of point functions with 0/1 outputs. 
Optimized FSS schemes for the offset class of useful gates, such as ReLU, splines (piecewise polynomials), and many more were presented in~\cite{boyle2019secure,EC:BCG+21,wagh2022pika,JawalkarGBCGS24,PoPETS:GJMCGPS24,SP:GCGKS25}.
See Section~\ref{sec:preprocessing} for further discussion of this application of FSS.

    
  \item {\bf NC$^0$ functions.}
    For {\em $d$-local} functions $f:\{0,1\}^n\to\{0,1\}^k$, in which each output bit depends on at most $d=O(1)$ input bits, a PRG-based DPF can be used to get FSS with key size $O_d(k\log n\cdot \lambda )$~\cite{EC:BoyGilIsh15} (and evaluation time that scales with $n^d$). For example, the case $k=1$ captures bit-matching predicates that depend on a constant number of input bits.
    This is based on the fact that FSS for an arbitrary function class of size $N$ reduces to a DPF with domain size $N$ by using the DPF to share the indicator vector of the function in the class. 
  
  \item {\bf Decision trees with topology leakage.}
  A decision tree is defined by: (1) a tree topology, (2) variable labels on each node $v$ (where the set of possible values of each variable is known), (3) value labels on each edge (the possible values of the originating variable), and (4) output labels on each leaf node.
  
In the construction of~\cite{CCS:BoyGilIsh16}, the key size is roughly $|V|\cdot \lambda$ bits, where $V$ is the set of nodes, and evaluation on a given input requires $|V|$ PRG invocations, and a comparable number of additions. The FSS is guaranteed to hide the secret edge value labels and leaf output labels, but (in order to achieve this efficiency) reveals the base tree topology and the identity of which variable is associated to each node.

\item {\bf Constant-dimensional intervals.}
A useful application of the above FSS scheme for decision trees is FSS for {\em $d$-dimensional interval functions}. Such a multi-dimensional interval is a $f(x_1,\ldots,\allowbreak x_d)$ which evaluate to a selected nonzero value precisely when $a_i \le x_i \le b_i$ for some secret interval ranges $(a_i,b_i)_{i \in [d]}$.
 For $n$-bit inputs $x_i$, FSS for $d$-dimensional intervals can be obtained with key size and computation time $O(n^d\cdot \lambda)$. For small values of $d$, such as $d = 2$ for supporting two-dimensional rectangles, this yields solutions with a
 reasonably good concrete efficiency. A lower bound of $\tilde\Omega(n^{1.5})$ on the evaluation time of FSS schemes for 2-dimensional intervals that make a {\em black-box} use of a PRG was recently obtained in~\cite{GW26} via a connection with dynamic data structures. 
  \end{itemize}

In the context of private information retrieval, FSS schemes for the above function classes can be combined with server-side database operations to emulate private database search with richer query classes, such as Max/Min and top-$k$~\cite{splinter}. 
See Section~\ref{sec:apps} for discussion of these and other applications of FSS from symmetric cryptography.

\medskip


\paragraph{Homomorphic Secret Sharing.}
When considering FSS for rich function classes, it is often more natural to use the dual notion of {\em homomorphic secret sharing} (HSS), where the roles of the input and the function are swapped. Namely, whereas FSS evaluates a secret-shared function on a public input, HSS evaluates a public function on a secret-shared input. Note that HSS and FSS are closely related, but that ``HSS for a function class $\mathcal{F}$'' is not identical to ``FSS for the class $\mathcal{F}$.''  Notably, HSS for $\mathcal{F}$ places a more stringent requirement on the share size of an input $x$ to grow only as a function of the size of $x$ and support homomorphic evaluation of functions $f \in \mathcal{F}$; in contrast, FSS for $\mathcal{F}$ must inherently have key size that grows with the description size of $f \in \mathcal{F}$.

FSS can be generically reduced to HSS by relying on universal functions. A universal function for a class $\cF$ is a function $U_\cF$ such that for all $f \in \cF$ and $x \in \zo^n$, it holds that $U_\cF(f,x) = f(x)$. Given such a universal function, one can reduce the problem of constructing FSS for $\cF$ to the problem of constructing HSS for the class of functions $\mathcal{U}_\cF := \{U_\cF(\cdot,x) : x \in \zo^n\}$. The key size of the resulting FSS scheme (inherently) scales with the description size of $f$.
\medskip

\subsubsection{``Mid Range'' -- FSS/HSS from LPN-Style Assumptions}

In recent years, there has been success in building lightweight constructions of HSS from variants of the Learning Parity with Noise (LPN)~\cite{BlumFKL93} assumption.  The LPN assumption asserts that it is computationally hard to solve a system of uniformly random linear equations over $\F_2$ where each bit is \emph{flipped} with some small probability $\epsilon$. 
It can be shown that this search variant of LPN is equivalent to the decision variant, asserting that noisy linear combinations are pseudorandom. 
An equivalent formulation of LPN is that it is hard to decode a noisy (random) codeword from a given random \emph{linear code}. Changing the distribution over equations is equivalent to considering other types of linear codes.
Many variants of LPN are standard in the literature, over fields other than $\F_2$, and with different distributions over linear codes and/or noise.

Depending on the parameter regime (e.g., noise rate), the LPN assumption and its variants can be viewed as lying in between symmetric cryptography and public-key cryptography. 
While LPN with low noise is known to imply public-key encryption~\cite{FOCS:Ale03}, it is not known to imply additively homomorphic encryption (or even collision-resistant hashing, except in an extreme parameter regime~\cite{BrakerskiLVW19}). 

In the context of HSS and especially in the related context of PCGs, LPN turns out to be surprisingly useful, even in a regime where it is not known to imply public-key encryption. Under LPN, the following HSS constructions are known:


\begin{itemize}
      \item {\bf HSS for constant-degree polynomials}, from standard LPN \cite{C:BCGIKS19}; 
      \item {\bf HSS for $\log\lambda/\log\log\lambda$-degree polynomials}, from the superpolynomial hardness of LPN~\cite{EC:CouMey21}. 
      \item {\bf Multi-party HSS for $\log\lambda/\log\log\lambda$-degree polynomials} from a sparse variant of LPN~\cite{DIJL23}. This construction applies to any number of parties $m$ with any linear secret sharing of the output. Combined with a linearly homomorphic encryption, it gives rise to a {\em homomorphic encryption scheme} for a similar class~\cite{CorriganGibbsHKV25}. On the downside, unlike previous LPN-based HSS schemes, this construction suffers from an inverse-polynomial error. 
      
      For the default notion of HSS, where the output shares are additive, the correctness error can be made negligible for the class of constant-degree polynomials~\cite{CouteauKY25}. This implies a multi-party DPF whose key size scales with $N^{1/d}$, for an arbitrary constant $d$. The construction relies on a combination of (non-sparse) LPN and a flavor of the MQ assumption, asserting that a random system of quadratic equations is hard to solve. This too qualifies as a ``mid-range'' construction, since the underlying assumptions are not known to imply public-key cryptography.
 \end{itemize}
 
In all of the above cases, the share size scales linearly with the input size and is sublinear in the description size of the function being evaluated. This rules out, for instance, HSS for constant-degree polynomials that uses additive sharing of all monomials (as was the case for the information-theoretic FSS for low-degree polynomials described above). An HSS scheme for degree-2 polynomials with compact (but not additive) output shares was constructed in~\cite{CatalanoF15} from threshold additively homomorphic encryption. Constructions of compact non-additive HSS for low-degree polynomials from partially homomorphic encryption schemes were given in~\cite{LaiMS18,IshaiLM21}.


\medskip


\subsubsection{``High End'' -- FSS/HSS from Public-Key Cryptography}



Using standard public-key cryptography assumptions, HSS schemes exist for much richer function representation classes, including branching programs (capturing functions in $\mathsf{NC}^1$ and logspace) and even general circuits (capturing general polynomial time computations). Concretely, the following results are known.

      \begin{itemize}
      \item {\bf Branching programs} (inverse-poly error): Allowing for inverse-polynomial error (namely,  $\delta$-correctness for any inverse polynomial $\delta$), a construction from Decisional Diffie-Hellman (DDH) was presented in~\cite{C:BoyGilIsh16}. 
      In this construction the running time of $\Eval$ is $\tilde O(s^2/\delta)$, where $s$ is the branching program size and $\delta$ is the error probability (which can be made detectable, namely of a Las-Vegas type). This was subsequently improved to $\tilde O(s^{1.5}/\delta^{0.5})$ in~\cite{C:DinKelKle18}. Optimized variants of the DDH-based constructions for simple but useful function classes are given in~\cite{EC:BoyGilIsh17,CCS:BCGIO17}. 
      A similar construction from the Decisional Composite Residuosity (DCR) assumption was presented in~\cite{FazioGJS17}.
      
      \item {\bf Branching programs} (negligible error): With full (negligible-error) correctness, HSS constructions from the DCR assumption were presented in~\cite{EC:OrlSchYak21,C:RoySin22}. In these constructions, the running time scales linearly with~$s$.

      \item {\bf Circuits}: A general-purpose FSS/HSS scheme for Boolean circuits based on the Learning With Errors (LWE) assumption was presented in~\cite{C:DHRW16} via a flavor of homomorphic encryption called ``spooky encryption.'' The construction combines a multi-key variant of fully homomorphic encryption with a local rounding procedure for converting noisy additive shares of the output over a big modulus to almost noiseless additive shares of the same output value over a small modulus. 
      
     Note that a more efficient construction of HSS from LWE (and Ring-LWE) was given for the case of branching programs by~\cite{EC:BKS19}; this construction avoids homomorphic multiplication of encrypted values and instead emulates restricted multiplications via a form of distributed decryption. Additionally, a more direct construction of FSS for branching programs from LWE (and a small exponent variant of the DCR assumption) was given by~\cite{AC:BKLS24}, bypassing the overhead of universal circuits inflicted by the HSS-to-FSS transformation.

      \end{itemize}

\section{Applications}
\label{sec:apps}

In this section, we discuss some of the main applications of DPF and, more generally, FSS schemes for simple function classes, to concretely efficient secure computation and more. 
We start (Section~\ref{sec:DistribGen}) with a distributed key generation protocol which is needed in several of these applications, and then move on to describe different classes of applications.

\subsection{Efficient Distributed Key Generation}
\label{sec:DistribGen}

In many applications of DPFs, there is no single entity who knows the identity of the secret point function; instead, the role of ``client'' is jointly executed across parties. In these cases, the point function itself is either generated or defined by values held secret shared across parties, and the $\Gen$ algorithm of the DPF must in turn be executed distributedly via a secure computation protocol.  
This is the situation, for example, for applications of DPFs to secure computation for RAM programs~\cite{doerner2017scaling} or mixed-mode operations~\cite{boyle2019secure,EC:BCG+21}, use of pseudorandom correlation generators for secure computation preprocessing~\cite{CCS:BoyleCGI18,boyle2019secure,boyle2020efficient}, and more, as we discuss in the sections below. 

Secure execution of $\Gen$ can of course always be achieved generically, by running an off-the-shelf secure computation protocol run on the corresponding $\Gen$ procedure expressed as a Boolean (or other) circuit. However, doing so requires unrolling the cryptographic pseudorandom generator (PRG) operations as complex Boolean circuits, which is typically undesirable for efficiency.
Alternatively, work has gone toward designing targeted secure protocols for achieving the same goal, while remaining {\em black box} in the underlying cryptographic PRG. In particular, the parties locally evaluate the PRG on various inputs, but will never need to use (or know) the explicit code or circuit implementation of the PRG itself.

This was initiated by the work of Doerner and shelat~\cite{doerner2017scaling}. In this work (using DPFs for efficient secure computation of RAM programs), they presented a distributed $\Gen$ protocol for the standard tree-based DPF construction~\cite{CCS:BoyGilIsh16}, which requires computation time linear in the DPF domain size $N$, and $\log N$ sequential communication rounds, but which crucially makes only {\em black-box} use of oblivious transfer and a pseudorandom generator. 
Their protocol remains the most efficient to date for this task. 

Note that in relaxed variants where one
party is allowed to learn $\alpha$ (but still not $\beta$), the protocol can be parallelized, reducing the round
complexity to constant~\cite{CCS:BCGIKR19,CCS:SchoppmannGR019}.

\paragraph{Black-box Distributed $\Gen$~\cite{doerner2017scaling}.}
A DPF key for the tree-based DPF is made up of an initial seed, which is randomly and independently sampled for each party, and $n+1$ public correction words. The $i$-th correction word, $1 \le i \le n$ is used to ensure that the values in the two children of $v$, the $i$-th node on the path from the root to $\alpha$ are suitable. That is, the values in the two trees of the child that is off-path must be identical, while the values of the child on the path are pseudorandom. That is achieved by setting the correction word to be the XOR of the two values of the off-path child, and having a single party XOR this value with its off-path child during $\Eval$. Denote the seed at the left child of $v$ by $s_{0,\ell}$ for the first party and by $s_{1,\ell}$ for the second party, and denote the seeds of the right children by $s_{0,r}, s_{1,r}$. Consider the expansion of all the seeds in the $i$-th level to strings of double the seed length, and divide them into all the left halves of these strings and all the right halves of these strings. \cite{doerner2017scaling} observes that if the values of all the nodes in the $i$-th level are identical in both trees except for the values of $v$, then the XOR of all the left halves in the first party is $L \oplus s_{0,\ell}$ and in the second party the XOR is $L \oplus s_{1,\ell}$, for some identical string $L$ that depends on all the nodes in the $i$-th level that are not on the path to $\alpha$. Similarly, the XOR of all the right halves are $R \oplus s_{0,r}$ and $R \oplus s_{1,r}$ for the first party and for the second party, respectively. It follows that the $i$-th correction word is the XOR of all the left halves in both trees or the XOR of all the right halves in both trees. Determining which option is correct depends on $\alpha$.

Given a bit by bit secret sharing of $\alpha$ and an additive secret sharing of $\beta$ in its target group $\bbG$, the distributed generation protocol of \cite{doerner2017scaling} begins with each party independently sampling a random initial seed. The parties compute the correction words in order from the first to the $n$-th. To compute the $i$-th correction word, the first party acts as the sender in an Oblivious Transfer (OT) protocol for strings, setting up two messages $m_0,m_1$. If its share of the $i$-th bit of $\alpha$ is $0$ then it assigns $m_0=L \oplus s_{0,\ell}$ and $m_1 = R \oplus s_{0,r}$. If its share of the $i$-th bit of $\alpha$ is $1$ then it switches between the values of the two strings. The second party acts as the OT receiver and learns $m_b$ for its share $b$ of the $i$-th bit of $\alpha$. If $b=0$ then the correction word is $m_b \oplus (L \oplus s_{1,\ell})$, while if $b=1$ then the correction word is $m_b \oplus (R \oplus s_{1,r})$. Therefore, the second party can compute the $i$-th correction word after the OT and share it with the first party. Computing the last correction word, which additively masks $\beta$ does not even require an oblivious transfer. Each party computes the expansion of a seed in all the leaves of the tree into elements in $\bbG$. A fixed linear combination of these values and of the shares of $\beta$ gives the last correction word. The number of symmetric key operations, i.e., PRG expansions, in this protocol is $O(2^n)$, while the number of string OTs is exactly $n$. 

We refer readers to Section 5 of~\cite{doerner2017scaling} for a detailed description and pseudocode of the corresponding Distributed $\Gen$ procedure.
This procedure was extended to handle distributed comparison functions in~\cite{EC:BCG+21}.

The main overheads of the protocol are the $O(N)$ computational cost for both parties, and the $\log N$ round complexity.
While there are no known ways to avoid the former without making non-black-box use of cryptographic primitives, the round complexity can in some cases be reduced.
If we are in a relaxed setting (as in some applications) where it is allowed to leak the secret index $\alpha$ to one of the two parties, the round complexity can be brought down to just two rounds, since all OTs can be performed in parallel~\cite{boyle2019secure,CCS:SchoppmannGR019,CCS:BCGIKR19}.
Another approach based on a so-called \emph{programmable} DPF~\cite{boyle2022programmable} can achieve a constant round complexity without leaking $\alpha$, although with poor concrete efficiency. 

In the setting of distributed multi-point functions, distributed setup protocols can get more involved, particularly for the more complex batch code or OKVS based DMPF constructions from Section~\ref{sec:DMPF}.
A recent work~\cite{ARR25} improves this by using a ``reverse cuckoo hashing'' based DMPF that is designed to simplify the setup protocol: instead of sampling hash functions and then obliviously assigning items into bins, the construction first picks a valid assignment and then solves a system of linear equations (inside 2-PC) to determine the hash functions.



\subsection{Private Reading}

In the introduction, we already sketched a simple application to the task of \emph{private information retrieval} (PIR), where a client holding a private keyword $\alpha$ can perform a lookup to a database $D$ held by two servers, such that the client learns whether $\alpha \in D$, by using a DPF where $\alpha$ is encoded into a point function $f_{\alpha,\beta}$ for $\beta=1$. This approach implies simple and fast PIR implementations; see~\cite{GPU} and references therein.

We now explore more expressive extensions of this.
Consider a more structured database $D$, consisting of key-value pairs $(v_i,x_i)$ instead of just keywords.
Using a DPF, we can extend the previous construction to perform a \emph{sum} query, where the client learns the sum (over the DPF output group $\G$) of all the $x_i$'s where $v_i = \alpha$.
This can be again done by generating DPF keys $k_0,k_1$ for the point function $f_{\alpha,1}$, and having each server $b$ compute
\[ y_b =
\sum_{(v_i,x_i)\in D}
\Eval(b,k_b,v_i) \cdot x_i. \]

After receiving each server's shares, the client can reconstruct the result $y_0 + y_1$.

Going beyond equality constraints, using a distributed comparison function (DCF) instead of DPF allows the client to select (or sum) values that fall in a given interval.
Various matching and aggregation functions of this type can be implemented using DPFs and DCFs, to allow for private database queries supporting a relatively large subset of SQL, as explored in~\cite{splinter}.


\paragraph{Unbalanced Private Set Intersection.}
Another use-case for this type of protocol is the task of private set intersection, where the client holds a small set $C$ and wishes to learn the intersection with the server's database $D$, which is now a simple set of keywords $\{v_i\}_i$.
By relaxing the standard PSI setting to a two-server model, where $D$ is held by two non-colluding servers, the parties can run the naive PIR lookup for every element $\alpha \in C$, allowing the client to learn $C \cap D$.
Note that the communication complexity of this protocol is independent of the size of $D$, unlike standard PSI solutions in the literature.
This makes it particularly suitable for scenarios such as private contact discovery, where the server's database is much larger than the client's set.

Further extensions and optimizations to this approach were considered in~\cite{PoPETS:DRRT18}.

A form of function secret sharing has also been used to obtain efficient {\em structure-aware} PSI, where one party's input set has a publicly known structure, yielding fuzzy PSI and beyond~\cite{C:GarRosSin22,C:GarRosSin23,C:GarGofMia24,AC:BGMP25}.



\subsection{Private Writing}
\label{sec:private-writing}

In a \emph{writing} analog of PIR, instead of reading, the client wishes to periodically perform updates to a database by writing secret values.
This was first considered by Ostrovsky and Shoup~\cite{ostrovsky1997private} in the information-theoretic setting, and solutions based on distributed point functions were proposed in~\cite{corrigan2015riposte,CCS:BoyGilIsh16}.
In this application, it is inherent that no single server can know the database, which is typically enforced by secret-sharing it among the servers.
Suppose that two servers hold additive shares $v_{0,i}$ and $v_{1,i}$ of each $i$-th entry in $D$, such that $v_i = v_{0,i} + v_{1,i}$.
A client who wishes to update entry $\alpha$ from $v_\alpha$ to $v_\alpha'$ can create DPF keys $k_0,k_1$ for the function $f_{\alpha,\beta}$, where $\beta = v_\alpha' - v_\alpha$.
If server $b$ then updates its shares to $v_{b,i}' = v_{b,i} + \Eval(k_b, i)$, the shared database is updated accordingly.

\paragraph{Private Aggregate Statistics.}
This type of private writing can be applied to the task of privately gathering aggregate statistics from users of complex software such as a web browser or operating system.
Such aggregate statistics are useful for identifying popular features, detecting malicious websites, tracking error rates, and more, while ensuring that the contribution of any individual user remains private.

As a concrete example (proposed in~\cite{CCS:BoyGilIsh16}), consider a \emph{private histogram} protocol. Suppose there is a set of possible websites, indexed by the set $S = [n]$, and a population of clients who each hold a private value $\alpha_i \in S$ (e.g. a website that triggered a browser crash), and the goal is to compute the histogram counting how many clients experienced a crash under each of the sites in $S$, without revealing any individual input.
Using the private writing technique above, each client generates DPF keys $(k_{i,0}, k_{i,1})$ for the point function $f_{\alpha_i,1}$ and sends $k_{i,b}$ to server $b$. Each server $b$ then computes, for every $j \in S$, the sum $h_{b,j} = \sum_i \Eval(k_{i,b}, j)$. By the correctness of the DPF, the combined histogram $h_{0,j} + h_{1,j}$ counts exactly the number of clients with input $\alpha_i = j$, while the security of the DPF ensures that no single server learns any individual client's input.

Prio~\cite{prio} was the first practical system to address this type of private aggregation at scale.
While Prio does not rely on DPFs, instead using standard secret-sharing, the authors noted that DPFs could be used to compress the share size in use-cases such as histograms.

The private histogram protocol above requires the set $S$ to be small enough for the servers to enumerate.
In many settings, however, the domain of possible inputs is exponentially large (e.g., arbitrary URLs or search queries), making full enumeration infeasible.
The problem of \emph{private heavy hitters} asks to identify the most frequently occurring strings among the client population, without learning any individual input and without enumerating the full domain. Boneh et al.~\cite{boneh2021lightweight} introduced the Poplar protocol, which solves this problem by relying on an \emph{incremental} variant of the tree-based DPF (Section~\ref{sec:construct}) that allows evaluation on any prefix of an input.
The key idea is that each client submits incremental DPF keys encoding their private string, and the servers use the incremental structure to efficiently search for heavy hitters by progressively expanding only the prefixes that appear frequently, rather than evaluating over the entire domain.

Doplar~\cite{PoPETS:DPRS23} builds on the Poplar approach, reducing the number of rounds of server interaction needed to verify that client submissions are well-formed.

\paragraph{Anonymous Messaging.}
The private writing paradigm can also be applied to build anonymous messaging systems.
The Riposte~\cite{corrigan2015riposte} system is an anonymous bulletin board, where a set of servers hold secret shares of the messages to be sent over a given time period.
A client who wishes to publish a message $m$ uses the private writing protocol to insert it at a random position in the bulletin board, sending a DPF key to each server to allow it to update its shares.
After all clients have submitted their keys in a given epoch, the servers jointly reconstruct the bulletin board to reveal the posted messages, while the DPF security ensures that no single server learns which client wrote to which row.

A key challenge in this setting is handling \emph{write collisions}: if two clients attempt to write to the same row, their messages will be summed together, corrupting both.
Riposte addresses this by increasing the database size to accommodate the expected number of writes, together with some algebraic error correction techniques to allow recovery in case of a small number of collisions.

\paragraph{Handling Malicious Clients with Verifiable DPFs.}
In the above applications, the servers must trust that each client's DPF keys encode a valid point function.
A malicious client could submit keys that do not correspond to a valid point function, or with a nonzero point outside of a required range, which could corrupt the aggregation results.
To mitigate this, many works add a \emph{verification} procedure that allows the servers to check that keys are well-formed, whilst preserving privacy of the hidden point against a semi-honest server.

One approach to supporting verifiability is via information-theoretic \emph{arithmetic sketching} techniques, first used in~\cite{CCS:BoyGilIsh16} and later developed in~\cite{C:BBCGI23}.
We give a simple example for proving that a DPF with domain size $N = \mathsf{poly}(\lambda)$, and outputs shared over a finite field $\F$ of characteristic $>2$, encodes a single non-zero value $\beta \in \{0,1\}$.
After receiving the DPF keys from the client, the servers will agree upon some common randomness (for instance, derived from a shared PRF key) $r_1, \dots, r_N \in \F$.
Let $(y_1, \dots, y_N)$ be the vector of all $N$ evaluations of the point function.
The idea is that the servers will compute shares of $z_1 = \sum y_i r_i$ and $z_2 = \sum y_i r_i^2$, and then use a small 2-PC protocol to check that $z_1^2 = z_2$.
This holds if and only if

\begin{align*}
     0 
        &= \left( \sum_{i=1}^N y_i r_i \right)^2 - \sum_{i=1}^N y_i r_i^2 \\
        &= \sum_{i=1}^N (y_i^2 - y_i) r_i^2 + \sum_{1 \le i < j \le N} 2 y_i y_j r_i r_j
\end{align*}
Viewing this as a degree-2 polynomial in $(r_1,\dots,r_N)$, notice that if there are two non-zero outputs $y_i$ and $y_j$ then the monomial $r_i r_j$ has a non-zero coefficient (since the characteristic is not two), and so by the Schwartz-Zippel lemma, the probability of passing the verification is at most $2/|\F|$.
Similarly, in case there is only a single non-zero output, $y_i$, but it lies outside $\{0,1\}$, then $y_i^2 - y_i \ne 0$, and again we can apply Schwartz-Zippel.

In~\cite{CCS:BoyGilIsh16,C:BBCGI23}, various other randomized sketching procedures are given for testing different kinds of relations, including different possibilities for $\beta$ other than $\{0,1\}$, supporting arbitrary finite fields, and supporting weight-$w$ vectors for verifying distributed multi-point functions.

Another approach to making a DPF verifiable is to use a cryptographic hash function.
This approach relies specifically on the tree structure of the construction from Section~\ref{sec:constructions}; it was initially used to verify consistency of a distributed puncturable PRF protocol in~\cite{CCS:BCGIKR19}, and extended to obtain a verifiable DPF in~\cite{EC:CastroP22}.
The idea is that the tree will be extended by one level, so that the number of DPF outputs grows from $N$ to $2N$. The extra $N$ values will all be zero, meaning that both servers obtain the same set of shares (after correcting for sign).
Thus, the servers can just hash together the $N$ extra shares, then exchange and compare the resulting hash values.
It was shown in~\cite{EC:CastroP22} that, under certain collision-resistance and correlation-intractability assumptions on the hash function, this guarantees consistency of the DPF outputs, even if one of the servers is malicious.

This approach has the benefit of a lower round complexity than arithmetic sketching approaches, since the servers do not need to agree upon any randomness, and furthermore, no general-purpose 2-PC is needed.
On the other hand, cryptographic hashing can be more computationally expensive than finite field operations, as observed in~\cite{PoPETS:DPRS23}.

Both approaches to achieve verifiable DPFs (arithmetic sketching and using cryptographic hash functions) suffer from computation that is linear in the input domain size. Circumventing this obstacle is possible in some applications by considering only a subset of the input domain and verifying the output of the DPF only on that subset. That approach was adopted in \cite{boneh2021lightweight} for the subset of {\em heavy hitters}, i.e., the most popular query strings that are supplied by multiple clients. 

However, limiting the consideration to a subset of the domain enables a subtle attack when the subset to be considered is unknown at $\Gen$ time. A malicious client can submit a query with multiple non-zero locations and will only be discovered if the subset of interest happens to include two or more non-zero inputs. In this way, a malicious client casts several ``votes'', gaining more influence on the final result than an honest client.

{\em Extractable DPF} is a refinement of standard DPF that blocks this type of double-voting attack. At a high level, the structure of the queries in extractable DPF ensures that a malicious client can only choose one point at which the sum of evaluations of the point has a {\em permissible}, non-zero value, where only a small subset of elements of the output group have a permissible value for a point function. In slightly more detail, a $2$-party extractable DPF scheme in the Random Oracle Model is equipped with a polynomial time algorithm, called the extractor, which given as input the keys that $\Gen$ constructs and all of the calls to the random oracle that $\Gen$ makes returns as output an input location $\alpha$. The adversary wins if $\Eval(0,k_0,\alpha^*)+\Eval(1,k_1,\alpha^*)=\beta$ for some $\alpha^* \ne \alpha$ and a non-zero $\beta$, which is permissible as the output of a point function. The scheme is secure if the probability that the adversary wins is negligible. 

In \cite{boneh2021lightweight}, it was proven that the DPF presented in Figure \ref{fig:DPFx4} is an extractable DPF, given two assumptions: the PRG is modeled as a Random Oracle, and the output group of the point function is sufficiently large. Several other DPF constructions are not extractable, including the less efficient $2$-party DPF schemes in \cite{EC:BoyGilIsh15,EC:GilboaIshai14} and the multi-party DPF schemes in \cite{EC:BoyGilIsh15,goel2025multiparty}. Intuitively, the DPF in Figure \ref{fig:DPFx4} is  extractable because a malicious client cannot control more than one non-zero path in the GGM tree.

In any level $i$ of the tree, an honest client uses the single correction word at its disposal to ensure that the evaluation of the function in one additional sub-tree, rooted at the off-path child of the $i$-th level on-path node, is zero. A malicious client has two possible strategies. First, it can generate a correction word that does not set the value in either child to zero. However, in this case, modeling the PRG as a random oracle implies that all of the two or more non-zero values are random and independent of each other, ensuring that the client can determine the final output of at most one leaf. If the output group is large enough in relation to the subset of permissible outputs of the point function, then the probability that the value at more than one point will be in the permissible set is negligible. A second strategy for a malicious client is to provide different correction words to the two servers. Indeed, each of the keys $k_0,k_1$ has a {\em private} part, which includes the initial PRG seeds used at the root of the tree, and a {\em public} part, which includes all the correction words. The private part is sampled independently for each key, while the public part is identical in an honest execution of $\Gen$. To test that the client acts honestly, providing identical correction words in each key, the two servers can run a secure protocol to ensure that the two public parts of their keys are identical.

To informally summarize the analysis on the DPF in Figure \ref{fig:DPFx4}: if the input domain is $\{0,1\}^n$, the output group is $\mathbb{G}$, the set of permissible values for a point function is $P \subseteq \mathbb{G}$, the PRG is modeled as a random oracle with a malicious client making at most $t$ calls to this PRG, and the two servers securely testing that the public parts of their keys are identical, then the DPF is extractable and the probability that a malicious client wins the security game is at most $\frac{O((t^2+nt)|P|)}{|\mathbb{G}|}$.

\subsection{Secure Computation with Preprocessing}
\label{sec:preprocessing}
\newcommand{\Gin}{\mathbb{G}_{in}}
\newcommand{\Gout}{\mathbb{G}_{out}}


A common paradigm for efficient secure two-party computation splits the protocol into an input-independent {\em preprocessing} (or offline) phase and a lightweight {\em online} phase. In the preprocessing phase, a trusted dealer (or an additional cryptographic protocol) distributes correlated randomness to the two parties. Once the inputs become available, the online phase uses this correlated randomness to evaluate the desired function with minimal interaction.

Boyle, Gilboa, and Ishai~\cite{boyle2019secure} introduced a simple and general approach for secure computation with preprocessing, building on function secret sharing. This approach can be viewed as a generalization of the ``TinyTable'' protocol of Damg{\aa}rd et al.~\cite{damgaard2017tinytable}: whereas TinyTable represents the correlated randomness for each gate as a full truth table (whose size is exponential in the number of input wires), the FSS-based approach replaces these truth tables with FSS keys, achieving exponential compression for gate types that admit efficient FSS schemes. The approach was further developed in~\cite{EC:BCG+21}, extending it to ``mixed-mode'' secure computation where different parts of the circuit operate over different algebraic domains.

\paragraph{Overview of the approach.}
The central observation is best illustrated for a single gate. Consider a gate $g : \Gin \to \Gout$ that two parties $P_0, P_1$ wish to evaluate on a secret input. Suppose that, rather than holding secret shares of the input $w$, both parties know a {\em masked} version $\hat{w} = w + \rin$ for a random mask $\rin \in \Gin$ unknown to either party individually. If the dealer knows $\rin$ as well as a random output mask $\rout \in \Gout$, it can precompute the {\em offset} function $g_{\rin,\rout}(\hat{w}) = g(\hat{w} - \rin) + \rout = g(w) + \rout$ and distribute FSS keys for this function to the two parties. During the online phase, both parties evaluate their FSS keys on the common value $\hat{w}$, obtaining additive shares of $g(w) + \rout$. After exchanging these shares, they recover the masked output $g(w) + \rout$, which is itself a fresh masking of $g(w)$, to use as input to the next gate.

To evaluate an entire circuit $C$, the dealer associates a random mask $r_j$ with every wire $j$ in $C$, and for each gate prepares FSS keys for the corresponding offset function as above. The protocol maintains the invariant that both parties learn the masked value $w_j + r_j$ for every wire $j$. At the input level, this is established by having each party send its masked input to the other; at each subsequent gate, the FSS-based procedure above advances the invariant from input wires to output wires. At the end, the dealer reveals the output-wire masks so the parties can recover the result.

The efficiency gain over TinyTable comes from the FSS compression: instead of storing a full truth table for each gate's offset function, the parties store only FSS keys. For gate types that admit compact FSS schemes, such as comparison and equality gates via DCF and DPF (Section~\ref{sec:constructions}), the correlated randomness per gate is exponentially smaller than the truth table.\footnote{A general method for compressing truth-table correlations was suggested in~\cite{C:BCGIKS19}. However, the {\em running time} still grows linearly with the truth-table size, or exponentially with the gate input length.}

A variant of this protocol, also described in~\cite{boyle2019secure}, makes the correlated randomness {\em circuit-independent} by choosing masks for each gate independently and providing additive shares of the input masks alongside the FSS keys. The parties can then non-interactively reconcile the mask mismatch between adjacent gates during the online phase, at the cost of slightly higher communication (one group element per wire rather than per wire value).

\paragraph{Instances and applications.}
Several well-known protocols in the preprocessing model can be viewed as special cases of this FSS-based framework. For gates computing a product over a ring (i.e., $g(x_1,x_2) = x_1 \cdot x_2$), the corresponding offset function class has an FSS scheme that reduces to the familiar {\em Beaver triple} technique~\cite{Beaver91a}. For Boolean gates on a small number of input bits, the truth-table-based FSS recovers the TinyTable protocol~\cite{damgaard2017tinytable}.

The approach becomes particularly powerful for gates with structured functionality that admit efficient FSS but have large truth tables. Using DPF-based and DCF-based FSS, the following types of nonlinear gates can be efficiently implemented with compact correlated randomness and optimal online communication~\cite{boyle2019secure,EC:BCG+21}:
\begin{itemize}
    \item {\em Equality test and integer comparison:} Given a masked arithmetic input, the parties can securely evaluate whether two values are equal, or which is larger, using DPF and DCF respectively.
    \item {\em Bit decomposition:} Converting an arithmetic sharing over a large group into sharings of the individual bits. This can be realized by composing comparison gates.
    \item {\em Fixed-point arithmetic:} Operations such as truncation and rounding, which are essential for fixed-point representations of real numbers, can be expressed using comparison-based gates.
    \item {\em Nonlinear activation functions:} Piecewise-linear or piecewise-polynomial approximations of functions such as ReLU or sigmoid can be evaluated using a small number of comparison and arithmetic gates.
\end{itemize}

These capabilities make the FSS-based preprocessing approach especially useful for privacy-preserving machine learning, where secure computation must handle a mix of linear operations (e.g., matrix multiplications over arithmetic shares) and nonlinear operations (e.g., activation functions and argmax). Works such as~\cite{EC:BCG+21} and subsequent systems~\cite{PoPETS:GJMCGPS24,SP:GCGKS25} have demonstrated that FSS-based techniques offer a competitive approach for such mixed-mode computations.

\subsection{Pseudorandom Correlation Generators}
\label{sec:PCG}

As well as the general FSS-based approach above, DPFs have been instrumental in building \emph{pseudorandom correlation generators} (PCG), which can produce large quantities of correlated randomness for use in MPC protocols in the preprocessing model with only a small amount of interaction~\cite{CCS:BoyleCGI18,C:BCGIKS19}.
A PCG has a key generation algorithm that outputs a pair of short, correlated seeds, which can then be locally (or, silently) expanded to obtain a large quantity of correlated randomness.
The correlated randomness might take the form of, for instance, a large batch of random oblivious transfer (OT) correlations, or a batch of Beaver triples.
We give an overview of some simple constructions of PCGs based on the \emph{learning parity with noise} assumption (LPN).

To start with, observe that a DPF is a natural way to compress pseudorandom secret shares of a \emph{sparse} vector.
A weight-one vector $\vec v = (0, \dots, \beta,0, \dots, 0) \in R^N$ over a ring $R$ can be viewed as the truth table of the point function $f_{\alpha,\beta} : [N] \rightarrow R$.
Hence, a dealer can distribute secret shares of $\vec v$ to two parties by generating and giving out a pair of DPF keys.
This easily extends to sharing weight-$t$ sparse vectors for not-too-large $t$, by using a multi-point DPF.

We can combine the compressed sparse vector sharing technique with LPN to obtain various forms of correlated randomness.
The dual form of LPN over $R$ states that for a secret, sparse vector $\vec e \in R^n$ and public, random matrix $H \in R^{m \times n}$, the product $H \vec e$ is computationally indistinguishable from a random vector (this is also known as the syndrome decoding problem).
We first apply this to generating a so-called \emph{vector oblivious linear evaluation} correlation (or vector-OLE), where one party holds vectors $\vec u, \vec v$ while another party holds a scalar $x$ and learns $\vec w = \vec u x + \vec v$.
The goal is for two parties to obtain a single, long instance of a random vector-OLE, given only a pair of short seeds.
Viewing vector-OLE as a (subtractive) sharing of $\vec u x$, we will define a pseudorandom $\vec u$ vector as an LPN instance $H \vec e$, and then use a multi-point DPF to give to the parties shares of $\vec e x$, which has the same sparsity as $\vec e$.
After expanding the DPF seeds, each party can then locally multiply their share vector by $H$ to obtain the respective $\vec v$ or $\vec w$ output.
As part of the PCG seeds, we additionally give one party $x$ and the other party the sparse $\vec e$ vector, which completes the correlation setup.

This PCG template can be extended beyond just a vector-OLE correlation.
Firstly, vector-OLE can be generalized to give \emph{correlated OT}, a form of OT where every pair of the sender's messages satisfies $m_{i,0} + m_{i,1} = \Delta$ for some fixed, secret offset $\Delta$. This can be shown to be equivalent to vector-OLE where the $u_i$ entries are restricted to $\{0,1\}$, also called \emph{subfield vector-OLE} when the ring $R$ is a binary extension field; furthermore, correlated OT can be converted into standard, random OT by putting the messages through a correlation-robust hash function~\cite{C:IKNP03}.

Secondly, we can also use DPFs to obtain a PCG that produces a large batch of OLE correlations of the form $w_i = u_ix_i + v_i$, where all the $u_i, v_i, x_i$'s are independently random.
This is done via a simple tensor product trick: sample two LPN error vectors $\vec e, \vec e' \in R^n$, and define $\vec u = H \vec e$ and $\vec x = H \vec e'$.
We have:
\[ \vec u \cdot \vec x^\top = H \cdot \vec e \cdot(\vec e')^\top H^\top \]

$\vec e \cdot (\vec e')^\top$ is a tensor product matrix with at most $t^2$ non-zero entries.
Therefore, it can be distributed into shares using a $t^2$-point DPF with domain size $N^2$.
After this, the parties can locally compute shares of $\vec u \cdot \vec x^\top$ and take the diagonal of this matrix to obtain shares of $u_i x_i$ for all $i$ (as a by-product, they can additionally compute shares of any bilinear function of $\vec u$ and $\vec x$).

\paragraph{Optimizations and Extensions.}
The above construction for vector-OLE was first introduced in~\cite{CCS:BoyleCGI18}, while those for OT and OLE were given in~\cite{C:BCGIKS19}. 
Since then, PCGs have become a highly active area of research, and many variants of the above constructions with improved efficiency characteristics exist in the literature.
Instead of using a uniform matrix $H$, practical constructions typically choose a structured matrix of a special form allowing for fast multiplication, thus relying on a structured variant of LPN instead of the standard LPN assumption.
For example, $H$ could be a quasi-circulant matrix~\cite{CCS:BCGIKR19}, a polynomial multiplication matrix (as in ring-LPN~\cite{boyle2020efficient}), or chosen as the parity-check matrix chosen from a carefully designed family of codes with fast encoding and good minimum distance~\cite{boyle2022correlated,raghuraman2023expand}.
Instead of the dual form of the LPN assumption given above, some works use LPN in its (standard) primal form $A\vec s + \vec e$, where in the \emph{sparse-LPN} variant, each row of $A$ is chosen to be sparse, significantly speeding up computation~\cite{YWLZW20}.

\paragraph{Pseudorandom Correlation Functions.} A useful extension of PCGs that enables a virtually unbounded number of correlated randomness instances is a {\em pseudorandom correlation function} (PCF)~\cite{boyle2020correlated}. While a standard PCG allows parties to expand short, correlated seeds into a fixed-size batch of correlated pseudorandom strings, a PCF takes this a step further by acting as the ``function'' analogue. This is analogous to the way a standard pseudorandom function (PRF) extends a standard PRG. A PCF for simple correlations such as OT can be obtained from any FSS scheme for a class of (weak) PRF. While an LWE-based feasibility follows from FSS for general circuits, this approach is impractical. The first concretely efficient PCF construction from~\cite{boyle2020correlated} relied on a new weak PRF candidate, in which a function in the class is a sum of point functions on variable-size subsets of the input. An FSS scheme for this class can thus be based on multiple DPF instances. Subsequent constructions of PCF for OT and other correlations were given in~\cite{boyle2022correlated,BuiCMPR24,BCMRS25}.





\subsection{Locally Random Reductions}
\label{sec:lrr}

We conclude with a complexity theoretic application of DPFs. Using a sub-exponentially secure DPF, which can in turn be based of a sub-exponentially-secure one-way function, we can obtain a (relaxed variant of) {\em locally random reduction}, for functions in high complexity classes that uses only two queries.
This, in turn, can be used to improve over classical results on worst-case to average-case reductions for these classes.

For context, classical arithmetization techniques in complexity theory~\cite{BF90,Lipton,BFNW} use polynomials to efficiently reduce the evaluation of $f(x)$ to the evaluation of a related function $g$ on inputs $x_1,\ldots,x_n$, where each input $x_i$ is {\em individually} random. Furthermore, if $f$ is in PSPACE or EXPTIME, then so is $g$. This is referred to as an $n$-query locally random reduction (LRR). Such an LRR can be used for the following kind of worst-case to average-case reduction: given an oracle to any algorithm $A_g$ for $g$ which is correct on all but an $\epsilon$-fraction of the inputs, we can efficiently implement a probabilistic algorithm $A_f$ for $f$ which is correct for {\em all} inputs, except with $\epsilon n$ error probability, by having $A_f(x)$ invoke $A_g$ on $x_1,\ldots,x_n$ obtained by the LRR. By a union bound, the probability that $A_g$ is incorrect on {\em any} of these inputs is at most $\epsilon n$, implying a similar upper bound on the error probability of~$A_f$.

As observed in~\cite{EC:GilboaIshai14}, a 2-party DPF can yield a similar LRR with only {\em 2 queries} $x_0,x_1$: On input $x$, the reduction invokes $\Gen(x)$ to obtains DPF keys $k_0,k_1$. The function $g$ is the DPF $\eval$ function. For existing DPF constructions, if $f$ is in PSPACE or EXPTIME, then so is $g$.

Note, however, that this requires settling for a weaker notion of LRR, where each query $x_i$ is {\em pseudorandom} rather than perfectly random. (Pseudorandomness of keys holds for standard DPF constructions, including those described in Section~\ref{sec:constructions}.) However, assuming a sub-exponentially secure PRG, which follows from a  one-way function with similar security, we can still ensure that if $A_g$ runs in (fixed) exponential time, the pseudorandomness of the DPF is good enough to fool $A_g$ while still allowing for polynomial-time key generation. 

This relaxed notion of LRR yields worst-case to average-case reduction as above for functions $f$ in PSPACE or EXPTIME with only 2 queries. This implies that if $A_g$ is correct on all but an $\epsilon$-fraction of the inputs, $A_f$ errs with at most $2\epsilon$ probability. 

An extension of this idea to efficiently checking the computation of polynomial-time computable functions was explored in~\cite{BGILT18}. However, this extension requires FSS or HSS schemes for more complex function classes, which can currently only be constructed from ``public-key'' cryptographic assumptions such as the LWE.

\section{Open Questions}
\label{sec:openQ}

We close with a selection of open problems in the area.

\begin{itemize}

    \item Improved 2-party DPF: Can the $\approx n\cdot \lambda$ key size of the best known PRG-based construction be improved?  Here even concrete efficiency improvements (e.g., improved multiplicative constant) would be highly motivated.

    Alternatively, is there a way to formalize a barrier toward such improvement, given only a restricted set of computational operations?

    \item Multi-party DPF: 
    Can one beat the square-root key-size barrier for PRG/OWF-based full-threshold multi-party DPF?  In particular, is there a 3-party DPF secure against two corruptions with key size $O(\lambda\cdot 2^{(1/2-\epsilon)n})$ for some constant $\epsilon>0$? A canonical milestone is $O(\lambda\cdot 2^{n/3})$, but any constant improvement in the exponent would be a major advance.

    \item FSS for conjunctions: Which cryptographic assumptions suffice for efficient FSS for the class of arbitrary {\em conjunctions} of input bits? Are one-way functions sufficient? Alternatively, does this imply public-key cryptography? Note that conjunctions of up to $d$ bits can be achieved but with key size scaling as $O(2^d)$. 
    
    \item Efficient distributed $\Gen$: Is there a distributed generation algorithm for DPF 
    which makes a black-box use of a PRG with computation time that is polylogarithmic, or even just sublinear, in the domain size? For reference, existing distributed $\Gen$ protocols are either non-black-box in the PRG (e.g., na\"{i}ve secure evaluation of the $\Gen$ circuit), or require linear computation time in the domain size~\cite{doerner2017scaling,boyle2022programmable}.
    
    \item Information-theoretic DPF: Is there a {\em perfectly} secure 3-party DPF with domain size $N$ and key size $N^{o(1)}$? Here we only consider security with threshold $t=1$, namely each individual key perfectly hides the secret point. Such a 4-party DPF, or alternatively a {\em statistically} secure 3-party DPF, was presented in~\cite{BGIK22}. 
    
    What is the minimal key size of an information-theoretic 2-party DPF with compact output shares (say, 2-bit long) if we allow a general (non-additive) reconstruction of the output from the shares? What can we say about the power of information-theoretic FSS in this setting?
\end{itemize}

\bibliographystyle{IEEEtran}
\bibliography{refs}

\begin{IEEEbiographynophoto}{Elette Boyle}
is a Senior Scientist in the Cryptography and Information Security (CIS) Laboratory at NTT Research and an Affiliate Professor at Reichman University, Israel. She received her Ph.D.\ at MIT and B.S.\ at Caltech, both in Mathematics, and served as a postdoctoral fellow at Technion Israel and Cornell University.
Her research centers in cryptographic solutions for safely maintaining and processing sensitive data. In particular, her recent focus has been on protocols for secure multi-party computation, as well as underlying primitives such as function and homomorphic secret sharing. Her work has been recognized by awards from the European Research Council (ERC), Israeli Science Foundation (ISF), United States Air Force Office of Scientific Research (AFOSR), Google Research Scholar Award program, and the International Association of Cryptologic Research (IACR).
\end{IEEEbiographynophoto}

\begin{IEEEbiographynophoto}{Niv Gilboa} is an Associate Professor of Computer Science at Ben-Gurion University, Israel. He received a Ph.D. from the Faculty of Computer Science in the Technion, and did postdoctoral work at BGU. His research interests are broadly in secure computation, with an emphasis on function secret sharing, pseudorandom correlation generators, and fully linear proof systems. His work has been funded by awards from the Israeli Science Foundation (ISF), the Israeli Ministry of Science and Technology (MOST) and the European Horizon program.

\end{IEEEbiographynophoto}

\begin{IEEEbiographynophoto}{Yuval Ishai}
is a Professor of Computer Science at the Technion, Israel, currently on a sabbatical at AWS, whose research spans cryptography and computational complexity theory. His works were recognized by best paper awards of the FOCS 2004, Crypto 2007, and Crypto 2016 conferences, a CCS 2025 distinguished paper award, a SIAM Outstanding Paper prize, and a TCC Test of Time Award. He is a fellow of the International Association for Cryptologic Research and served as a program chair of the TCC 2011, Eurocrypt 2019, and Eurocrypt 2020 conferences.
\end{IEEEbiographynophoto}

\begin{IEEEbiographynophoto}{Peter Scholl}
is an Associate Professor in the Cryptography \& Cyber Security group at Aarhus University. He has worked extensively on developing efficient cryptographic protocols for secure multi-party computation, correlated randomness generation, zero-knowledge proofs and post-quantum cryptography. His work has been funded through a Sapere Aude award from the Independent Research Fund Denmark (DFF), as well as various projects with DARPA, the European Union and partners in industry.
\end{IEEEbiographynophoto}

\end{document}